\documentclass[fleqn,usenatbib]{mnras}

\usepackage{newtxtext,newtxmath}

\usepackage[T1]{fontenc}

\DeclareRobustCommand{\VAN}[3]{#2}
\let\VANthebibliography\thebibliography
\def\thebibliography{\DeclareRobustCommand{\VAN}[3]{##3}\VANthebibliography}

\usepackage{graphicx}	
\usepackage{amsmath}	
\usepackage{siunitx}

\DeclareSIUnit\parsec{pc}
\DeclareSIUnit\ph{ph}
\DeclareSIUnit\year{yr}
\DeclareSIUnit\simsun{M_\odot}
\DeclareSIUnit\ergs{ergs}
\DeclareSIUnit\ev{eV}
\DeclareSIUnit\byte{B}
\DeclareSIPrefix\comovmega{cM}{6}
\DeclareSIPrefix\comovkilo{ck}{3}
\DeclareSIUnit\h{h}
\DeclareSIUnit\angs{\textup{\AA}}
\DeclareSIUnit\atom{H}
\DeclareSIUnit\zsun{Z_\odot}

\newcommand{\rtwo}{$\rm r_{200}$}
\newcommand{\magy}{$\rm M_{AB1600}$}
\newcommand{\magyint}{$\rm M^{int}_{AB1600}$}
\newcommand{\magyext}{$\rm M^{ext}_{AB1600}$}
\newcommand{\extinct}{$\rm A_{AB1600}$}

\newcommand{\dustier}{{\fontfamily{pnc}\selectfont {DUSTiER}}}

\newcommand{\coda}{{\fontfamily{pnc}\selectfont {CoDa}}}
\newcommand{\codai}{{\fontfamily{pnc}\selectfont {CoDa I}}}
\newcommand{\codaii}{{\fontfamily{pnc}\selectfont {CoDa II}}}
\newcommand{\codaiii}{{\fontfamily{pnc}\selectfont {CoDa III}}}

\newcommand{\fescd}{$\rm f^{dust}_{esc}$}
\newcommand{\fescg}{$\rm f^{gas}_{esc}$}
\newcommand{\fesct}{$\rm f^{gxd}_{esc}$}
\newcommand{\fesc}{$\rm f_{esc}$}
\newcommand{\mathang}{\text{ \normalfont\AA}}

\newcommand{\msun}{\rm M_{\odot}}

\title[DUSTiER: DUST in the Epoch of Reionization]{DUSTiER (DUST in the Epoch of Reionization): dusty galaxies in cosmological radiation-hydrodynamical simulations of the Epoch of Reionization with RAMSES-CUDATON}

\author[Joseph S. W. Lewis]{
Joseph S. W. Lewis$^{1,2,3}$,
Pierre Ocvirk$^{1}$,
Yohan Dubois$^{4}$,
Dominique Aubert$^{1}$,\newauthor
Jonathan Chardin$^{1}$,
Nicolas Gillet$^{1}$,
\'Emilie Th\'elie$^{1}$
\\
$^{1}$Observatoire Astronomique de Strasbourg, Université de Strasbourg, CNRS UMR 7550, 11 rue de l’Université, 67000 Strasbourg, France\\
$^{2}$Zentrum f\"ur Astronomie der Universit\"at Heidelberg, Institut f\"ur Theoretische Astrophysik, Albert-Ueberle-Stra\ss e 2, 69120 Heidelberg, Germany\\
$^{3}$Max-Planck-Institut f\"ur Astronomie, Konigst\"uhl 17, D-69117 Heidelberg, Germany\\
$^{4}$Institut d’Astrophysique de Paris, UMR 7095, CNRS, UPMC Univ. Paris VI, 98 bis boulevard Arago, 75014 Paris, France}

\date{Accepted 30/12/2022. Received 22/12/2022; in original form 3/11/2021}

\pubyear{2022}

\begin{document}
\label{firstpage}
\pagerange{\pageref{firstpage}--\pageref{lastpage}}
\maketitle

\begin{abstract}
    In recent years, interstellar dust has become a crucial topic in the study of the high redshift Universe. Evidence points to the existence of large dust masses in massive star forming galaxies already during the Epoch of Reionization, potentially affecting the escape of ionising photons into the intergalactic medium. Moreover, correctly estimating dust extinction at UV wavelengths is essential for precise ultra-violet luminosity function (UVLF) prediction and interpretation. In this paper, we investigate the impact of dust on the observed properties of high redshift galaxies, and cosmic reionization. To this end, we couple a physical model for dust production to the fully coupled radiation-hydrodynamics cosmological simulation code RAMSES-CUDATON, and perform a \SI{16}{\comovmega\parsec\cubed\per\h\cubed}, $2048^3$, simulation, that we call \dustier{} for DUST in the Epoch of Reionization. It yields galaxies with dust masses and UV slopes roughly compatible with constraints at z $\geq 5$. We find that extinction has a dramatic impact on the bright end of the UVLF, even as early as $\rm z=8$, and our dusty UVLFs are in better agreement with observations than dust-less UVLFs. The fraction of obscured star formation rises up to 45\% at $\rm z=5$, consistent with some of the latest results from ALMA. Finally, we find that dust reduces the escape of ionising photons from galaxies more massive than \SI{e10}{\simsun} (brighter than $\approx$ -18 \magy) by >10\%, and possibly up to 80-90\% for our most massive galaxies. Nevertheless, we find that the ionising escape fraction is first and foremost set by neutral Hydrogen in galaxies, as the latter produces transmissions up to 100 times smaller than through dust alone.
    
\end{abstract}




\section{Introduction}
Over the coming decade, a number of new observatories targeting Reionization will see first light. For instance, the JWST, Euclid, and the {Nancy Grace Roman} telescopes will greatly bolster high redshift galaxy catalogues, whilst allowing the detection of further and fainter galaxies than ever before. At the same times, radio astronomy experiments such as {the Hydrogen Epoch of Reionization Array}, or the Square Kilometer Array are set to usher in a new era of Reionization science with direct detection of the neutral Hydrogen gas in the intergalactic medium (IGM) during the Epoch of Reionization (EoR). This new observational capability will be a particularly useful new window into the early Universe, and of great interest for the study of the Epoch of Reionization. However, these advances require progress in our understanding of how the observational signatures that interest us are produced. This means establishing methods to extract astrophysical information from the new data, but also striving to better comprehend the complex underlying physical processes during Reionization.

With this latter goal in mind, much effort has been made to produce large scale cosmological simulations that also follow the hydrodynamics of gas and the radiative transfer of ionising photons \citep[e.g.][and \cite{dayal_early_2018} for a review of numerical simulations
in galaxy formation and the Epoch of Reionization]{pawlik_aurora_2017, ocvirk_cosmic_2016, ocvirk_cosmic_2020, rosdahl_sphinx_2018, ma_simulating_2018, trebitsch_obelisk_2020, kannan_introducing_2021, katz_non-equilibrium_2022}. Due to the inhomogenous nature of the reionization process, and to the small scales at which star formation and crucial feedback mechanisms occur, this ideally requires both large scales ($\approx 100$ comoving Mpc \cite{iliev_simulating_2014} to achieve useful 21cm predictions) and high physical resolution (ideally <10pc to at least resolve large molecular clouds). However this is computationally extremely challenging, and past and current studies must choose to either focus on the very large scales, required to provide useful 21cm predictions, while others focus on very high resolution inside galaxies, at the cost of volume and representativity.

The Cosmic Dawn (\coda ) simulations\footnote{\codai \, \citep{ocvirk_cosmic_2016}, and \codaii \, \citep{ocvirk_cosmic_2020}, but also \codai \, AMR \cite{aubert_inhomogeneous_2018}} are cosmological RHD simulations of galaxy formation during the EoR. \codai \, and \codaii \, were run using the RAMSES-CUDATON \citep{ocvirk_cosmic_2016} code. One of RAMSES-CUDATON's important highlights is the performance of it's radiative transfer module, owed to the code's hybrid CPU/GPU design. This allows the simulations to use a full speed of light (as opposed to a reduced, dual, or variable setup [as in \citet{katz_interpreting_2017}]; refer to \citet{gnedin_proper_2016,ocvirk_impact_2019,deparis_impact_2019} for a discussion on the impact of such methods). The CoDa project lies at an intermediate point in scale and resolution when compared to other simulation projects. Its simulations encompass large volumes (94.4$^3$cMpc$^3$ in \codaii \,), but do not resolve the ISM of star forming galaxies (physical resolution of 3.3 pkpc at z=6 in \codaii \,). \codaii \, is in good agreement with observational constraints on Reionization; such as the high-redshift ultra-violet luminosity function (UVLF) from \citet{bouwens_uv_2015, bouwens_z_2017}. The resolution and scale of CoDa makes it an ideal tool for investigating Reionization and Reionization effects over large scales, with a large, significant sample of galaxies. For example, \citet{dawoodbhoy_suppression_2018} have examined the suppression of star formation in low mass galaxies due to local Reionization, and \citet{lewis_galactic_2020} investigated the ionising photon budget of galaxies in \codaii \,. Also, the scale-resolution trade-off of \codaii \, makes it a very useful simulation to study Lyman-$\alpha$ radiative transfer through a reionising Universe \cite{gronke_lyman-alpha_2020,park_crucial_2021}.

 Despite these successes, the CoDa simulations over-estimate the post overlap average ionisation of the IGM and average ionising photon density. The possible reasons for this are many, and some of them are discussed in \citet{ocvirk_cosmic_2016,ocvirk_cosmic_2020,ocvirk_lyman-alpha_2021}. One possible explanation we set ourselves to investigate and quantify in this paper is dust, which is not accounted for in \codai \, nor \codaii \,. Indeed, it is increasingly clear that massive star forming galaxies in the high redshift Universe already contain large dust masses (significant fractions of their stellar mass, as in \citet{schaerer_new_2015,bethermin_evolution_2015,laporte_dust_2017,burgarella_observational_2020, dayal_alma_2022}). Dust is coupled to gas and ionising photons in several ways that can interest us. First and foremost, dust can act as an absorber of Lyman continuum (hereinafter LyC) photons (i.e. ionising). Since dust accumulates faster in more massive galaxies, it could disfavour the role of massive galaxies in reionising the Universe, and therefore affect the ionisation of the IGM after Reionization, since dusty galaxies will be weaker ionising sources. At the same time, accounting for dust and dust extinction in simulations is a necessary step towards reproducing UV extinction, and realistic UV luminosity functions. Finally, reddening of the UV continuum of galaxies due to dust can alter the slope of their UV continua, providing an additional constraint on the dust content of high redshift galaxies \citep{bouwens_census_2014}. As a consequence, simulations such as  \cite{wilkins_properties_2017,wu_photometric_2020,vijayan_first_2020,lovell_first_2021} have begun to investigate the extinction of the UVLF and reddening of the UV continuum of galaxies in the high redshift reionising Universe. At the same time, highly resolved simulations have been used to explore dust physics and their effects in smaller, more detailed volumes \citep[e.g.][]{trebitsch_obelisk_2020}. However, as of yet, there have been very few attempts \citep[][]{kannan_introducing_2021} to study the effects of dust on the process of Reionization itself, in a large cosmological volume and in particular in a fully coupled radiation-hydrodynamical framework. {Moreover, most of the aforementioned studies do not attempt to directly connect dust masses to extinction and re-processing, and rely instead on calibrated scaling relations based on metal or gas column density.}

In the present paper, we set out to prepare the next large scale Cosmic Dawn simulation (Cosmic Dawn III or \codaiii), by implementing the physical model for dust formation of Dubois et al. (in prep) within RAMSES-CUDATON, which we calibrate and use to take a first look at the possible effects of dust on Reionization. To study the impact of dust, we performed a $2048^3$, \SI{16}{\comovmega\parsec\cubed\per\h\cubed} simulation with our new version of RAMSES-CUDATON, that we called \dustier{} for DUST in the Epoch of Reionization.

In this paper, we first present the simulation code and setup in Sec. \ref{sec:rc}, we then move on to validate our dust model and its setup in \ref{sec:dust_control}. Then, we comment on the effects of dust. First, by determining the effects of dust extinction on our UVLF and on the fraction of obscured star formation. Second, we investigate the impact of dust and the escape of ionising photons from galaxies in Sec. \ref{sec:pred}. Finally, we summarise our findings in Sec. \ref{sec:concl}.

\section{Methodology}
\label{sec:mthd}

\subsection{Deployment and setup: presenting \dustier}
\label{sec:dustier}
\dustier{} is a new 2048$^3$, $16^3$\SI{}{\comovmega\parsec\cubed\per\h\cubed} cosmological radiation and hydrodynamics simulation aimed at studying the importance of dust in reionization studies. \dustier{} ran using the RAMSES-CUDATON code \citep[][]{ocvirk_cosmic_2016}. RAMSES-CUDATON results from the coupling between the cosmological galaxy formation simulation code RAMSES \citep{teyssier_cosmological_2002} and the ionising radiative transfer module ATON \citep{aubert_radiative_2008}. As such, it is a fully coupled radiation-hydrodynamics code, and has been used in a number of publications by our group, in particular the Cosmic Dawn simulations \citep{ocvirk_cosmic_2016,ocvirk_cosmic_2020}, and more recently \cite{ocvirk_lyman-alpha_2021} (hereinafter O21). More details about the simulation code can be found in Sec. \ref{sec:rc}. 

Table \ref{tab:codanew} gives an overview of the \dustier{} simulation setup.

\begin{table}
\centering
\begin{tabular}{l l}

\hline
\multicolumn{2}{c}{Cosmology} \\
\hline
$\Omega_\Lambda$ & 0.693 \\
$\rm \Omega_m$ & 0.307 \\
$\rm \Omega_b$ & 0.048 \\
$\rm H_0$ & \SI{67.77}{\kilo\meter\per\second\per\mega\parsec} \\
$\rm \sigma_s$ & 0.8288 \\
n & 0.963 \\
$\rm z_{start}$ & 50\\
$\rm z_{end}$ & 4.5\\
\hline
\multicolumn{2}{c}{Resolution} \\
\hline
 
Grid size & 2048$^3$\\
Comoving box size & \SI{23.61}{\comovmega\parsec} (\SI{16}{\comovmega\parsec\per\h})\\
Comoving force resolution & \SI{11.53}{\comovkilo\parsec}\\
Physical force resolution at $\rm z= $6 & \SI{1.65}{\kilo\parsec}\\

Dark matter particle number & $2048^3$ \\
Dark matter particle mass & \SI{5.09e4}{\simsun}\\

Stellar particle mass & \SI{11732}{\simsun} \\

\hline
\multicolumn{2}{c}{Star formation and feedback} \\
\hline
 
Density threshold for star formation & 50<$\rm \rho_{gas}$>\\
Temperature threshold for star formation & \SI{2e4}{\kelvin}\\
Star formation efficiency $\rm \epsilon_\star$ & 0.03\\
Massive star lifetime & \SI{10}{\mega\year}\\
Supernova energy & \SI{e51}{\ergs}\\ 
Supernova mass fraction, $\rm \eta_{SN}$ & 0.2 \\
Supernova ejecta metal mass fraction  & 0.05\\

\hline
\multicolumn{2}{c}{Dust model} \\
\hline

 $\rm f_{cond}$ & 0.001 \\
 max(DTM) & 0.5\\

\hline
\multicolumn{2}{c}{Radiation} \\
\hline

Stellar ionising emissivity model & BPASS V2.2.1 binary \\
{\scriptsize  \citep[from][]{eldridge_population_2020}}\\
Stellar particle sub-grid escape fraction $\rm f_{esc}^{sub}$ & 1 \\

Effective photon energy & \SI{20.28}{\ev}\\
Effective HI cross-section (at \SI{20.28}{\ev}) & \SI{2.493e-22}{\meter\squared}\\

Dust mass attenuation coefficient values$^\dagger$ :\\
{\scriptsize\citep[SMC \& LMC values from][]{draine_infrared_2001}}\\

\qquad$\rm \kappa_{d, 611\mathang}^{SMC}$ (RT run) & \SI{8.85}{\m\squared\per\gram}\\
\qquad$\rm \kappa_{d, 611\mathang}^{LMC}$ (post-process) & \SI{13.58}{\m\squared\per\gram} \\
\qquad$\rm \kappa_{d, 1500\mathang}$ & \SI{4.89}{\m\squared\per\gram}\\
\qquad$\rm \kappa_{d, 1600\mathang}$ & \SI{4.15}{\m\squared\per\gram}\\
\qquad$\rm \kappa_{d, 2500\mathang}$ & \SI{2.61}{\m\squared\per\gram}\\
\hline

\end{tabular}
\caption{Some of the essential parameters of the \dustier{} simulation. $\dagger$: the 611 \AA{}  wavelength corresponds to the effective energy of the ionising photon group we follow. The 1600 \AA{} wavelength is relevant for UV AB magnitude calculations. The 1500  and $2500$ \AA{}  wavelengths are the centers of the blue and red pseudo-filters we use to compute the UV slopes of our simulated galaxies. {In our run of the \dustier{} simulation, we made use of SMC dust mass attenuation coefficients, that gave the best calibration results at time. Since then, we have refined our post-processing, and find a better match using LMC $\rm \kappa$. Since overall we find that whatever the choice of extinction law, the effect of dust on Reionization is small, we prefer introducing this inconsistency which allows for more realistic reddening and extinction in galaxies.} For more details on the choice of the extinction curve $\rm \kappa_d$, refer to App. \ref{app:kappa_pick}.}
\label{tab:codanew}
\end{table}

Below, we present the core features of the code, along with the new implementation of the dust model, how extinction is handled and the setup of the new simulation \dustier{}.

\subsection{RAMSES with dust}
\label{sec:rc}

RAMSES \citep{teyssier_cosmological_2002} is a eulerian simulation code for hydrodynamics, N-body dark matter dynamics and star formation, that is very broadly used in the astrophysical community and well suited to high performance computing in massively parallel setups.

\subsubsection{N-body dynamics and hydrodynamics}
In RAMSES, collision-less dark matter and stellar particle dynamics are handled using a particle mesh integrator.
Gas dynamics are solved on a eulerian grid, using a second-order unsplit Godunov scheme \citep{teyssier_kinematic_2006,fromang_high_2006}  based  on  the  HLLC  Riemann solver \citep{toro_restoration_1994}. A perfect gas Equation of State (hereafter EoS) with $\gamma$= 5/3 is assumed. For more details, please refer to \citet{teyssier_cosmological_2002}.

\subsubsection{Star formation}
Star formation in RAMSES is implemented via a phenomenological description (that reproduces the power law found by \cite{kennicutt_star_1998}), first described in RAMSES in \citet{rasera_history_2006}. Stars are depicted as particles that represent entire stellar populations. The creation of stellar particles is allowed in cells that are dense enough ($\rm \rho_{\rm gas} > 50  \langle\rho_{\rm gas}\rangle$), at a rate $\dot{\rho_\star}$ dictated by the gas density $\rho_{\rm gas}$, free-fall time $\rm t_{\rm ff}$, and an efficiency parameter $\rm \epsilon_\star=0.03$ (following Eq. \ref{eq:SF}). Moreover, star formation is only allowed in cells that are cooler than $\rm T_{SF}=2.10^{4} \, K$. We found in O21 that this temperature criterion, in conjunction with higher resolution than in \codai \,\& \codaii, produced a more realistic Reionization, in particular post-overlap, and we therefore adopt this same sub-grid model for star formation, and calibrate it as in O21,  with a star formation efficiency of $\rm \epsilon_\star=0.03$. The star formation rate density in a cell can then be summarised with the following law:

\begin{equation}
    \dot{\rho_\star} = \epsilon_\star  \frac{\rho_{\rm gas}}{t_{\rm ff}}, \, \rm if \rho_{gas}>  50 \langle\rho_{\rm gas}\rangle \, \rm and \, \rm T<2.10^4 \, K  \,,
\label{eq:SF}    
\end{equation}
where  $\rm t_{\rm ff} = \sqrt{\frac{3 \pi}{32 G \rho_{gas}}}$ is the gas free fall time.

In cells where the density and temperature criterion are met, stellar particles are drawn from a poissonian distribution of masses that depends on the cells' densities. The minimum stellar mass is therefore chosen to be a small fraction of the baryonic mass resolution ($\rm M_\star^{\rm birth}=11732 \, \msun$). The mass of stellar particles depends on the cell gas densities, but is always a multiple of this mass. When stellar particles are formed, they are assigned the metallicity of the gas in their birth cell.

\subsubsection{Stellar feedback and Chemical enrichment}

When a stellar particle reaches an age of 10 Myr, $\rm \eta_{SN}=20$\% of it's mass is assumed to explode as supernovas. Each supernova event injects $10^{51}$ ergs of energy for every $10 \, \msun$ of progenitor into it's host cell, using the kinetic feedback of \citet{dubois_supernova_2008}. After the supernova event a long lived particle of mass $\rm M_\star = (1-\eta_{SN})M^{birth}_\star$ remains.

We use standard RAMSES \citep{teyssier_cosmological_2002} chemical enrichment. Supernova events eject metals into the host cell, which can then be advected as a passive scalar along with the gas. The mass fraction of ejecta in metals is $\rm y=0.075$, the remainder of the gas is ejected with the metallicity the stellar particle was assigned upon creation. The $\rm y$ and $\rm \eta_{SN}$ parameters were adjusted to values that are compatible with our stellar evolution model, and that best matched the predictions for the metallicity of high redshift galaxies.

\subsubsection{Dust model}

The biggest novelty in this paper, with respect to previous implementations and deployments of RAMSES-CUDATON, is our new implementation of a physical dust model, taken from Dubois et al. (in prep, see \cite{trebitsch_obelisk_2020} for a similar implementation in RAMSES). The main goal of the dust model is to provide a realistic dust mass in each cell, which we can then use to compute the extinction of star light. Since it is coupled to an already rather heavy simulation code, we mean to keep it as simple as possible, and for instance, we only consider a single dust grain size of \SI{0.1}{\micro\meter}, and assume a standard solar chemical composition.
The model tracks dust creation and destruction on the fly in all the cells of the computational domain, through several processes.

\paragraph{Dust production}
Dust is released by supernova explosions. A fraction $\rm f_{cond}$(dust condensation fraction) of the released metal mass condenses into dust grains when a stellar particle undergoes a supernova event. The dust mass in a cell is also increased by the accretion of gas phase metals onto existing dust grains (or dust grain growth), as follows \citep[][]{dwek_evolution_1998}:

\begin{equation}
 \dot{\rm M}_{\rm d}= \rm \Big(1-\frac{M_d}{M_{metal}}\Big) \frac{M_d}{t_{growth}} \, ,
\label{eq:dust_accr}
\end{equation}
where $\rm M_d$ is the dust mass, $\dot{\rm M}_{\rm d}$ its time derivative, $\rm M_{metal}$ is the total (gas and dust) metal mass, and $\rm t_{growth}$ is the growth timescale.

\begin{equation}
\rm t_{growth}= 100 \, \alpha^{-1}(T) \, a_{0.1} \, n_{gas}^{-1} \Big(\frac{T}{20K}\Big)^{-0.5} \, Myr \, ,
\label{eq:tim_dust_accr}
\end{equation}
with $\rm a_{0.1}$ the dust grain size (a representative value of \SI{0.1}{\micro\meter} is chosen), $\rm n_{gas}$ the gas density in \SI{}{\gram\per\cm\cubed}, and T the gas temperature in Kelvin. The dimensionless sticking coefficient of gas particles onto dust is denoted $\rm\alpha(T)$. In our application, its value is 1 up to T=\SI{e5}{\kelvin}, and follows $\rm \alpha(T)=(1+10 \big(\frac{T}{10^6K}\big))^{-1}$ as in \cite{novak_radiative_2012}, meaning sticking becomes less efficient as temperature increases.

\paragraph{Dust destruction}
Dust is destroyed by the shock waves of supernovas (inertial sputtering) following \eqref{eq:sne_dust_dest} :

\begin{equation}
\rm \Delta M_{dest,SN} = 0.3 \frac{M_{s,100}}{M_g}M_d \, M_\odot \, ,
\label{eq:sne_dust_dest}
\end{equation}
where $\rm M_g$ is the cell gas mass and $\rm M_d$ is the cell dust mass. $\rm M_{s,100} = 6800 E_{SN,51} M_\odot$ is an estimate of the mass of gas shocked at velocities larger than \SI{100}{\kilo\meter\per\second} obtained from the Sedov solution in a medium of homogeneous density, and $\rm E_{SN,51}$ is the SN explosion energy normalised by \SI{e51}{\ergs} \citep[][]{mckee_photoionization_1989}.

Thermal sputtering also destroys dust. One of the first accurate calculations of the rate of thermal sputtering was given by \cite{draine_destruction_1979}. Here we use a fit to the characteristic time of destruction by thermal sputtering taken from \cite{novak_radiative_2012}, whereby destruction by thermal sputtering becomes more efficient a high temperatures :

\begin{equation}
\rm t_{dest,sput} = 0.1 \, a_{0.1}  \, n_{gas}^{-1} \, \Big(1+\left(\frac{10^6K}{T}\right)^3\Big) \, Myr\, .
\label{eq:dust_ther_dest}
\end{equation}

Finally, we define the dust to metals ratio (DTM) as the fraction of metals in the form of dust grains:

\begin{equation}
    \rm DTM = \frac{M_{\rm dust}}{M_{\rm dust} + M_{\rm metals}}
\end{equation}
where $\rm M_{\rm dust}$ is the mass of dust, and $\rm M_{\rm metals}$ is the mass of metals in the gaseous phase. By definition $\rm 0\leq DTM \leq 1$. At the end of each hydrodynamical RAMSES time step, each cell's DTM is checked against a maximum parameter to avoid potentially turning all the metal mass into dust. 
The model's free parameters, $\rm f_{cond}$=0.001 and max(DTM)=0.5 were calibrated so as to reproduce observable constraints and comparable results to semi-analytical models from the literature. {In practice, max(DTM) was chosen as 0.5 so as to not overstep the detected upper limits on dust masses in high redshift massive star forming galaxies. At the same time, this restricts the average DTM of a galaxy to values comparable to the Milky Way ($\approx 0.44$), which one could reasonably expect to be a rough upper limit on the dust mass of its high redshift progenitors. For lower stellar mass haloes, where $\rm f_{cond}$ is important (see Sec. \ref{sec:dust_gals}), we chose a very low value of 0.001 for $\rm f_{cond}$. Effectively, this places a strong upper limit on the role of dust in faint galaxies, and limits dust's importance very early on in the simulation, giving a steep evolution of the cosmic dust density compatible with SAM findings. We further discuss these choices in \ref{sec:dust_control}.}

\subsection{ATON}

ATON \citep{aubert_radiative_2008} is a radiative transfer code based on the M1 closure for the Eddington tensor \citep{levermore_relating_1984}. It is coupled with models for Hydrogen ionisation-chemistry and photo-heating.

\subsubsection{Source model}

In most of our previous work using RAMSES-CUDATON (\codai \, \& \codaii), stellar particles were assigned a fixed ionising emissivity that was cut off after the massive stars of the stellar populations underwent supernova events. Our new approach simply updates the particles' emissivities following the BPASSV2.2.1 stellar population model \citep{eldridge_population_2020}. In particular, we compute the emissivity in the ionising band used for radiative transfer, as well as in two UV bands used in post processing to determine the photometric properties of galaxies (See Sec. \ref{sec:extc}).
However, as in \cite{ocvirk_lyman-alpha_2021}, our stellar particles have masses close to $10^4$ $\msun$ or more, i.e. a star cluster of intermediate mass. Such a cluster does not form its stars instantaneously, but over the course of a few Myr \citep[][]{hollyhead2015,wall2020}. To account for this, we proceed as in \cite{ocvirk_lyman-alpha_2021} and model the stellar particle as a population of constant star formation rate over 5 Myr, a timescale compatible with star cluster models of \cite{He2019,He2020} and compute the corresponding time-metallicity-dependent H-ionising and continuum emissivities using the adopted BPASS models. We also compute the effective photon energy, average and effective ionisation cross-sections for ionising photons given in Tab. \ref{tab:codanew}, following \cite{rosdahl_ramses-rt:_2013} Eqs. B3-B5, adopting for this an average absolute metallicity Z=$10^{-3}$ and integrating overs stars up to 10 Myr of age, after which the ionising emissivity becomes too small to impact the radiative parameters significantly.

Finally, thanks to the porting of ATON to cuda for NVIDIA GPUs \citep{aubert_reionization_2010}, hence CUDATON, resulting in a massive speedup of the radiative transfer module, we are able to use the full speed of light in this study, circumventing the need for reduced or variable speed of light approaches (refer to \citet{gnedin_proper_2016,ocvirk_impact_2019,deparis_impact_2019} for a discussion on the impact of such methods).
However, due to the aggressive optimisation of CUDATON, which requires simple, uni-grid computational domains, the adaptive mesh refinement (AMR) of RAMSES must be turned off. Therefore the simulations discussed in this paper have a unique grid of fixed resolution. 

\subsubsection{Hydrogen thermo-chemistry}
In order to self-consistently follow the ionisation state of Hydrogen gas, ATON computes the rate of photo-ionisation, collisional ionisation and recombination, and the resulting ionising photon consumption. In this work, we only consider the gas heating and cooling processes associated with Hydrogen. The gas internal energy changes are followed as explained in \cite{aubert_radiative_2008}, using the Hydrogen cooling and heating rates of  \citet{hui_equation_1997,maselli_crash_2003}.

\subsubsection{LyC radiative transfer through dust}
\label{sec:dust_rt}
In order to account for dust absorption during the ATON radiative transfer time steps as well as for our post-processing, we consider the dust optical depth:

\begin{equation}
\rm \tau_d=\rm \rho_d  \, \kappa_{d,611\mathang} \,  dx \, ,
\label{eq:dust_tau}
\end{equation}
where $\rm \rho_d$ is the dust mass density in a cell in g/cm$^3$, dx is a cell width in cm, and $\rm \kappa_{d,611\mathang}$ is the dust mass attenuation coefficient at $\rm 20.28 eV$ (or \SI{611}{\angs}), i.e. for our ionising photon group, in $\rm cm^2/g$.
{The \dustier{} simulation was ran with} an extinction curve derived by \citet{draine_infrared_2001} for the Small Magellanic Cloud (SMC), giving $\rm \kappa_{d,611\mathang} = \,$\SI{8.85}{\cm\squared\per\gram}. This is a fairly standard choice among dust models of the early Universe. It is motivated by the fact that extinction curves as detailed as those available for the {SMC or the LMC} are not available in the high-redshift universe, and the fact that {these objects are dwarf galaxies}, it is often considered an adequate choice, or more likely a makeshift approximation for the bulk of high redshift dwarf galaxies we simulate here, even though differences in dust compositions and size distributions between SMC/LMC and high-redshift galaxies are rather likely. {An investigation of the impact of using a LMC versus SMC extinction curve on reddening, extinction and the ionizing escape of photons is provided in App. \ref{app:dust_rt}.}

\subsection{Initial Conditions}

Initial conditions were generated using the code mpgrafic \citep{prunet_mpgrafic_2013}, producing density and velocity fields for dark matter and baryons at an initial redshift $ z_0\approx150$, for the following cosmology: $\rm \Omega_\Lambda=0.693, \, \Omega_m=0.307, \, \Omega_b=0.048, \, H0=67.77 (km/s)/Mpc, \, \sigma_8= 0.8288, \, n=0.963$, compatible with  \cite{planck_collaboration_planck_2018}.

\subsection{Halo detection, and galaxy definition}

To detect dark matter haloes throughout our simulations, we use the PHEW code that is directly built into RAMSES \citep{bleuler_phew_2015}. PHEW is based on a watershed algorithm, and we use the following setup for our cosmological simulations : saddle threshold= 200, peak to saddle ratio= 3, minimum mass= 200 particles.
The use of PHEW is very advantageous in our case as we do not need to post process our numerous calibration simulations to detect haloes. Since PHEW runs simultaneously with the simulation code and shares RAMSES' structure, it also has a relatively low performance cost.  

We assume galaxies to reside in haloes within a spherical boundary centred on the halo detection centre, and with a radius \rtwo \, (as a proxy for the virial radius). This has the advantage of allowing direct comparison with previous work within the CoDa project (see \cite{ocvirk_cosmic_2016,dawoodbhoy_suppression_2018,ocvirk_cosmic_2020,lewis_galactic_2020}). Again, in line with our previous work, we assume that each halo hosts a single galaxy which is valid in the majority of cases. The limitations of this definition are discussed to some extent for the \codaii \, simulation in the appendix of \cite{ocvirk_cosmic_2020}.

\subsection{Computing Extinction and reddening for simulated galaxies}
\label{sec:extc}

To compute the extinction at $\rm 1600 \,\text{\AA}$ (to study its impact on UV LFs), and the reddening of the UV continuum of galaxy spectra, we rely on a simple line of sight (LoS) based method. First we pick an observation point at an infinite distance from our haloes (in practice this means that our LoS follow one of the axes of the Cartesian simulation grid). Then for each halo, and for every star forming cell per halo we compute the optical depth at the relevant wavelengths between the cell centre and out to $2\times$\rtwo \, along the LoS, with the appropriate dust coefficient $\rm \kappa_d$ from Table \ref{tab:codanew}. We allow ourselves to stop our dust opacity integration at $2\times$\rtwo \, because the dust content in the IGM is very low, and because it is very unlikely that a LoS should cross another dust enriched galaxy. Using this method, our results are susceptible to LoS effects (i.e. the geometry of galaxies) just like observations.
In the rest of the paper, we discuss both the magnitude accounting for extinction by dust (\magyext) and the intrinsic magnitude (with no dust extinction: \magyint).

To quantify the reddening of the UV continuum of galaxies due to dust, we compute the slope of the UV continuum ($\beta$) of our simulated galaxies. 
In order to measure the slope $\beta$ one assumes, following \cite{calzetti_dust_1994}, that the galactic spectrum is well represented by a power-law between 1250 and 2600 \AA{}, so that $\rm f_\lambda \propto \lambda^\beta$, where $\beta$ is the slope of the galactic UV continuum, and $\rm f_\lambda$ is the flux in the galactic UV continuum at the wavelength $\lambda$. Therefore it is customary to fit the simulated spectrum with a power-law, which yields the slope $\beta$. Here we proceed slightly differently, using 2 pseudo-filters, one blue and one red in the UV range considered. The blue (red) pseudo-filter is centered at 1500 \AA{} (2500 \AA{} respectively). Both pseudo-filters are purely top-hat (i.e. transmission is 0 or 1) and have a full width of 400 \AA{}. For a given stellar population represented by a collection of star particles in the simulation, it is straightforward to sum their flux (dust-extincted or intrinsic depending on the focus of the section) through the pseudo-filters using our BPASS models, yielding F$_{1500}$ and F$_{2500}$, the blue and red UV fluxes. From this the UV slope $\beta $ is obtained as:
\begin{equation}
\rm \beta = \frac{log_{10} (F_{2500}/F_{1500})}{log_{10}(2500/1500)} \,
\label{eq:beta}
\end{equation}

Whenever magnitudes or UV slopes are computed for our simulated galaxies, we do not account for nebular emission lines {or continuum}. We show in Appendix \ref{app:emlines}, using a set of nebular emission lines pre-computed for BPASS, that their impact on either magnitudes or UV slopes is negligible in the 1000-$2500\text{\r{A}}$ range considered here and does not impact our conclusions. {Finally, although we do not currently have a model in place for nebular continuum emission, we remind the reader that it could have a substantial impact on our $\beta$ predictions. Indeed, \citep[][]{wilkins_photon_2016} show that modeling the nebular continuum emission could have a reddening effect, potentially boosting the $\beta$ of \dustier{} galaxies by $\rm \approx 0.1-0.3$.}

\section{Results: the dust in \dustier{}}
\label{sec:constr}

Here we examine the realism of dust in our simulation, when compared to the few available observational constraints, and results from semi-analytical models and simulations. First, we examine our predictions for the dust masses of galaxies to confirm the setup of our model for dust production. Then, we investigate our predictions for the reddening of the slope of the UV continuum of galaxies by dust to validate our model for the extinction and reddening of UV light by dust grains. Finally, we assess the impact of our modelling on the UVLF and the escape of ionizing photons to the IGM.

\subsection{The build up of dust}
\label{sec:dust_control}
\paragraph*{Cosmic dust}

\begin{figure}
    \includegraphics[trim={0cm 0cm 0.cm 0cm},clip,width=\columnwidth]{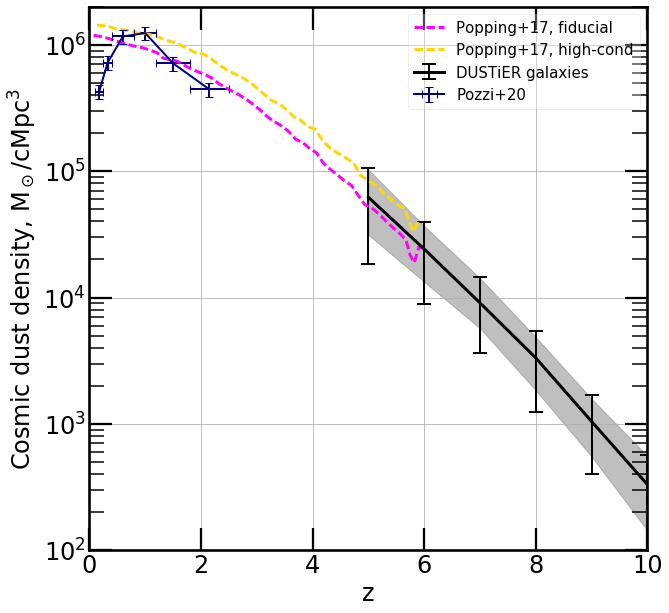}
    \caption{Cosmic dust density in \si{\simsun\per\comovmega\parsec\cubed} in \dustier{} (black full line), and the SAMs of \protect \cite{popping_dust_2017}. The blue data points are lower redshift constraints from \protect \cite{pozzi_dust_2020}. To produce the \dustier{} curve we summed the dust masses of all detected haloes at each redshift. To estimate the spatial variance in our result, we divided our volume into 8 equal cubic sub-volumes of \SI{8}{\comovmega\parsec\per\h} per side. The black error-bars represent the standard deviation of the total dust density across these sub-volumes, and the grey area shows the corresponding inter-quartile region.}
    \label{fig:cosmic_dust}
  \end{figure}

Fig. \ref{fig:cosmic_dust} shows the evolution of the box wide total average dust density with redshift. As one might expect based on the progressive build up of stellar mass in galaxies, the enrichment of galactic gas in metals and dust by successive stellar generations, as well as accretion onto existing dust grains, the total dust mass in our simulation rises with time. This is also the case in the semi analytical models of \cite{popping_dust_2017}, and in the simulations of \cite{graziani_assembly_2020}. We find that the build up of dust between $\rm z= 6$ and $\rm z= 5$ in our simulation agrees well with the predictions from the models of \cite{popping_dust_2017}. {We also include observational constraints from \cite{pozzi_dust_2020}, which seem roughly consistent with the evolution of the cosmic dust density in \dustier{}.} The total dust density can vary by up to a factor of 2 between 8 sub-volumes taken from our simulation, this spatial variance is greater than the difference between the two presented models from \cite{popping_dust_2017}.

\paragraph*{Dust in galaxies}
\label{sec:dust_gals}

  \begin{figure*}
    \includegraphics[trim={0cm 0cm 0.cm 0cm},clip,width=\columnwidth]{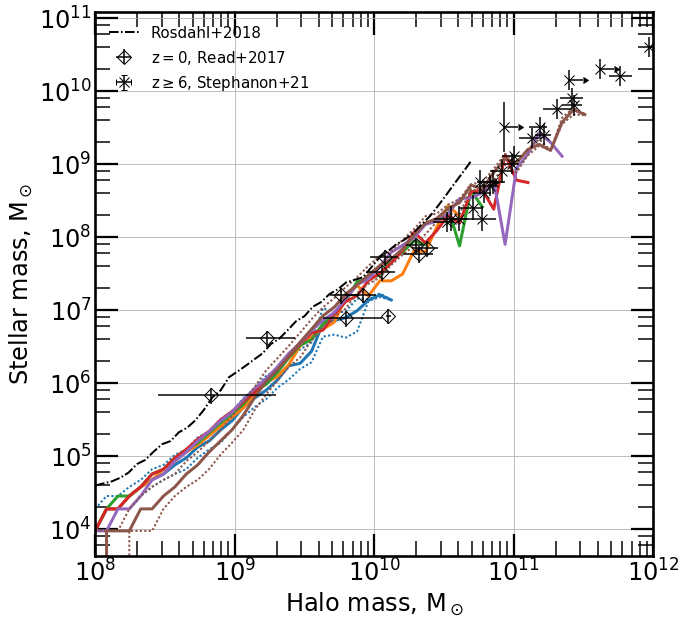}
    \includegraphics[trim={0cm 0cm 0.cm 0cm},clip,width=\columnwidth]{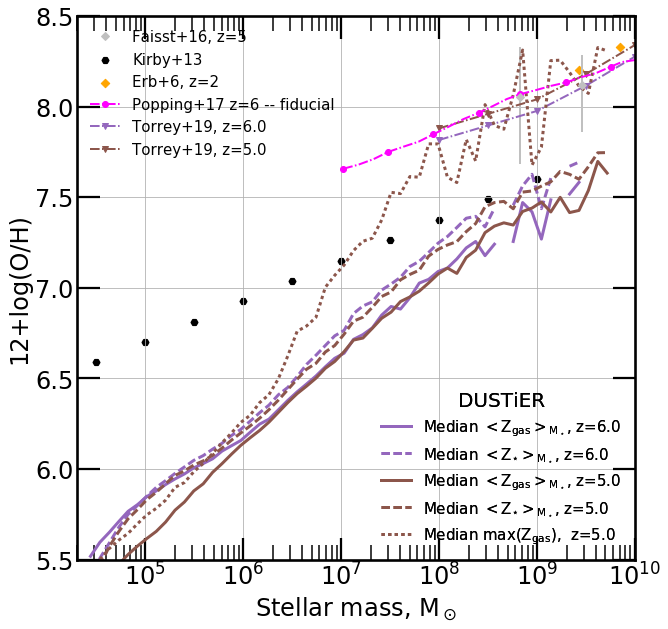}    
    \caption{{\emph{Left:}Median stellar mass to halo mass relation in \protect \dustier{}, compared to low redshift observations \protect \cite{read_stellar_2017}, high redshift observations from \protect \cite{stefanon_galaxy_2021}, and the SPHINX simulations \protect \citep[][]{rosdahl_sphinx_2018}. Dotted lines show the 84$\rm^{th}$ and 16$\rm^{th}$ percentiles for the first and last plotted redshifts. \emph{Right:} Metallicity statistics for \dustier{} galaxies. Also shown are the maximum values at $\rm z=6,5$. Where we have converted metallicities from Fe to O abundances using solar abundances. For comparison, constraints are plotted from \protect \citep[][]{erb_mass-metallicity_2006, kirby_universal_2013, faisst_rest-uv_2016}. The numerical results of \protect \cite{popping_dust_2017} and \protect \cite{torrey_evolution_2019} are also shown.}}
    \label{fig:gals_SMZ}
  \end{figure*}

{We first present some median properties of galaxies in Fig. \ref{fig:gals_SMZ} upon which our further study of dust will rely. The left hand panel shows the median stellar mass versus halo mass at several redshifts. The stellar mass of \dustier{} galaxies closely follows a powerlaw of halo mass, which shows close to no redshift evolution for $\rm z \geq 5$\footnote{Between $\rm z=6$, and $\rm z=5$, the median stellar mass decreases for the lowest mass galaxies ($\rm M_{halo}<$\SI{e9}{\simsun}). Based on prior work \citep[see][]{ocvirk_cosmic_2020,ocvirk_lyman-alpha_2021}, this is most likely a manifestation of star formation suppression brought about by Reionization. }. \dustier{}'s stellar mass to halo mass ratio is consistent with observations of high redshift galaxies in \cite{stefanon_galaxy_2021}, as well as with very high resolution radiation-hydrodynamical simulations such as SPHINX \citep[][]{rosdahl_sphinx_2018}.  Many galaxy properties in this paper are given as a function of stellar mass to emphasise the observational perspective when relevant. When needed, the reader may use this tight halo mass - stellar mass relation (HMSMR hereinafter) to convert to halo mass. This is mostly valid though above the mass scale where radiative suppression sets in, as otherwise the scatter of the HMSMR may increase.

The right hand panel shows the median mass weighted stellar metallicty versus stellar mass for \dustier{} galaxies at the same redshifts. For most stellar masses ($>$\SI{e6}{\simsun}), the metallicity increases with stellar mass, roughly following a powerlaw. This is expected when there is continuous star formation (with no suppression), as our initial mass function and supernova yields are fixed. We find that the typical metallicities from the literature have a similar slope, but lie roughly 0.5 dex above the equivalent galaxies in \dustier{}. However, we note the observational constraints carry large error bars, and that the most metal rich galaxies in \dustier{} are consistent with constraints. \cite{kirby_universal_2013} study the metallicities of local dwarf irregular galaxies, and report a gentler trend with stellar mass, but metallicities that are closer to \dustier{}'s for $\rm M_{stellar}\gtrsim$\SI{e8}{\simsun}.} 

\begin{figure}
    \includegraphics[trim={0cm 0cm 0.cm 0cm},clip,width=\columnwidth]{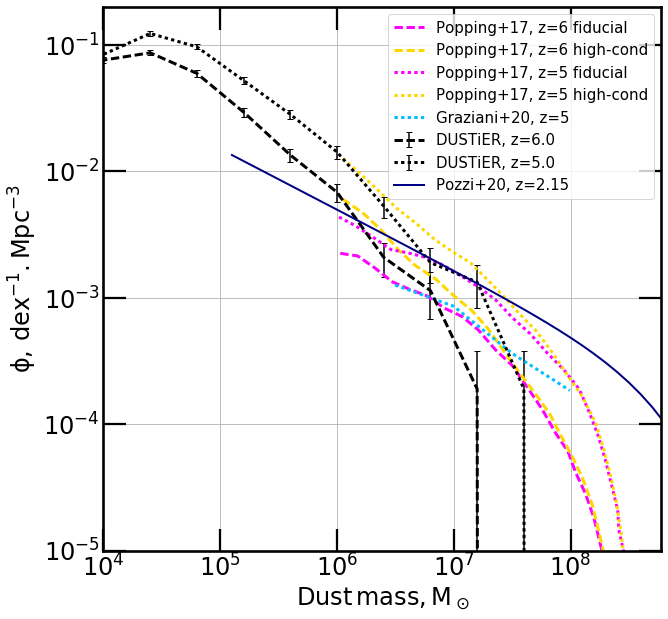}
    \caption{The dust mass function, or the number density of galaxies as a function of their dust mass. Black results are from our simulation. The black error bars represent the poissonian error on the DMF for each dust mass bin in \dustier{}. The yellow and magenta curves are the 'high-cond' and 'fiducial' from the SAMs of \protect \cite{popping_dust_2017}; and the blue line is from \protect \cite{graziani_assembly_2020}. The blue data points are lower redshift constraints from \protect \cite{pozzi_dust_2020}.}
    \label{fig:DMF}
  \end{figure}

We now move to study the dust properties of \dustier{} galaxies. The total mass of dust that forms in our simulation seems reasonable when compared to the existing literature. However, we can also compare our work in terms of the dust mass function (DMF), to check that the population of dusty galaxies is similar. Fig. \ref{fig:DMF} shows the DMF in our simulation. Broadly, the hierarchical nature of galaxy formation is imparted onto the DMF: the galaxies with the most dust are the rarest and the galaxies with the least dust are the most abundant. Over time more and more massive galaxies form and these can host higher and higher dust masses, and the normalisation of the DMF increases.
Here again, we find a good match to the literature: at $\rm z=5, 6$ our agreement with the 'high-cond' model of \cite{popping_dust_2017} is good near \SI{e6}{\simsun}. However at higher masses we under-predict the abundance high dust mass systems. This is in part due to the relatively small box size of \dustier{}, resulting in a lack of very massive haloes, causing the high dust mass cutoffs in the \dustier{} DMFs. Taking this into account (aided by the error bars that represent the poissonian error on the DMF within a mass bin) the agreement with the 'high-cond' model of \cite{popping_dust_2017} is fairly good for masses smaller than \SI{e7}{\simsun}, although their fiducial model, and also \cite{graziani_assembly_2020} show that the actual slope of the DMF could also be less steep and is not well constrained at high redshift. {The $\rm z=2.15$ DMF from \cite{pozzi_dust_2020} presents a much gentler slope than \dustier{}. At the high dust mass end this can be readily explained by the gradual build up of higher dust masses by $\rm z=2.15$, and by the modest box size of \dustier{}. For $\rm M_d \lesssim 10^6 M_\odot$, \dustier{} has an excess ($\rm \lesssim 0.5 \, dex \, at \, z=5$) of dust masses when compared to observations at $\rm z=2.15$. This could be the sign of too many small galaxies with too high dust mass to stellar mass ratios. However, it could also be partially explained by the hierarchical build-up of very massive dusty galaxies over time, driven by mergers of the least massive dusty galaxies in \dustier{}. Thus, an emptying of the low dust mass end, and a filling of the high dust mass end of the DMF could occur over time.}


\begin{figure*}
    \includegraphics[trim={0cm 0cm 0.cm 0cm},clip,width=0.95\columnwidth]{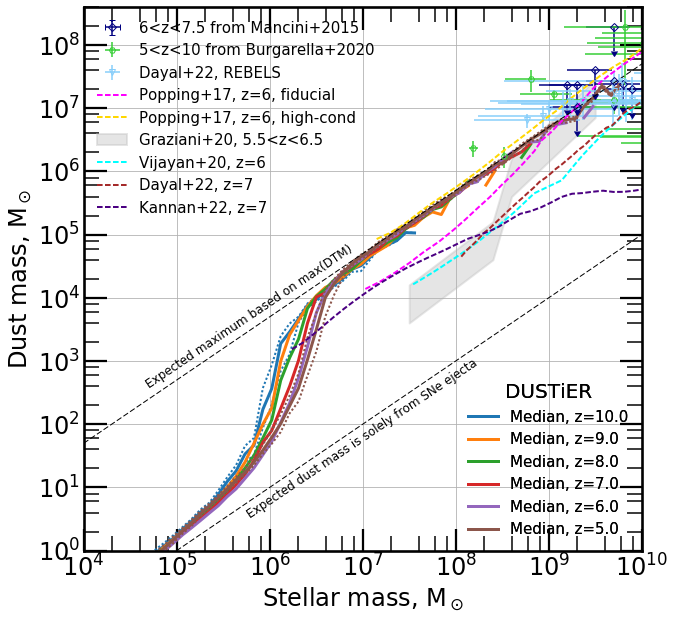}
    \includegraphics[trim={0cm 0cm 0.cm 0cm},clip,width=\columnwidth]{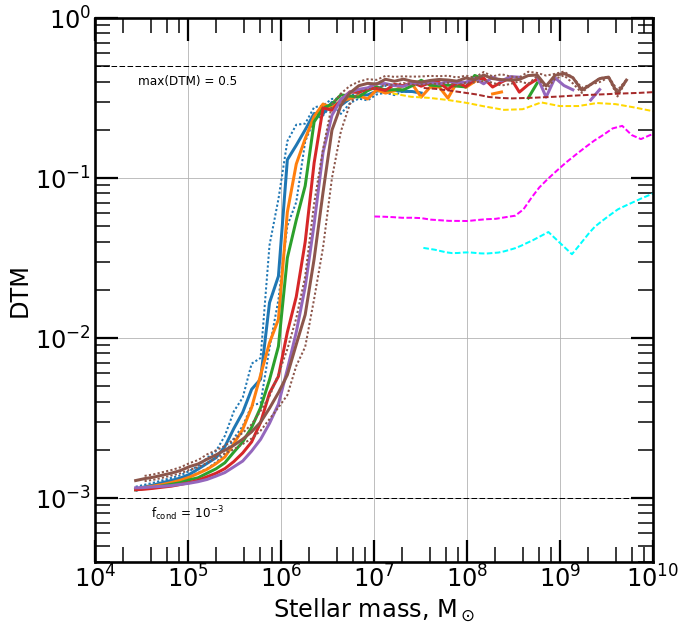}\\
    \includegraphics[trim={0cm 0cm 0.cm 0cm},clip,width=\columnwidth]{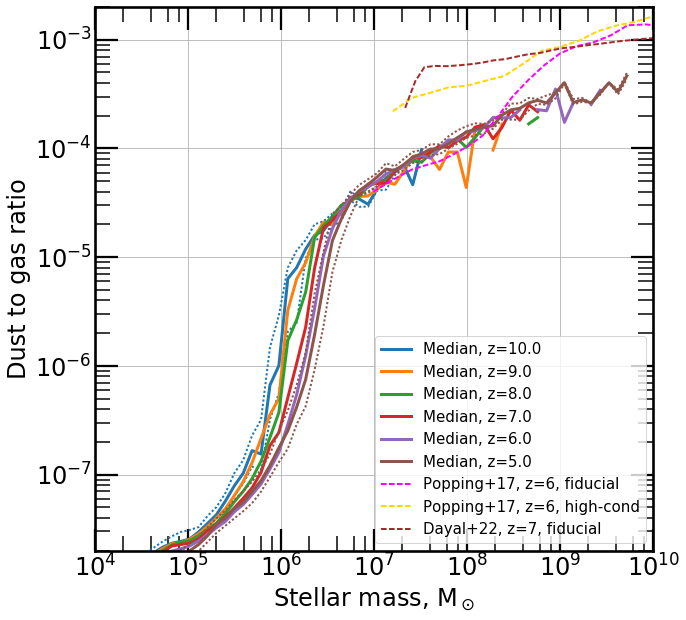}
    \includegraphics[trim={0cm 0cm 0.cm 0cm},clip,width=\columnwidth]{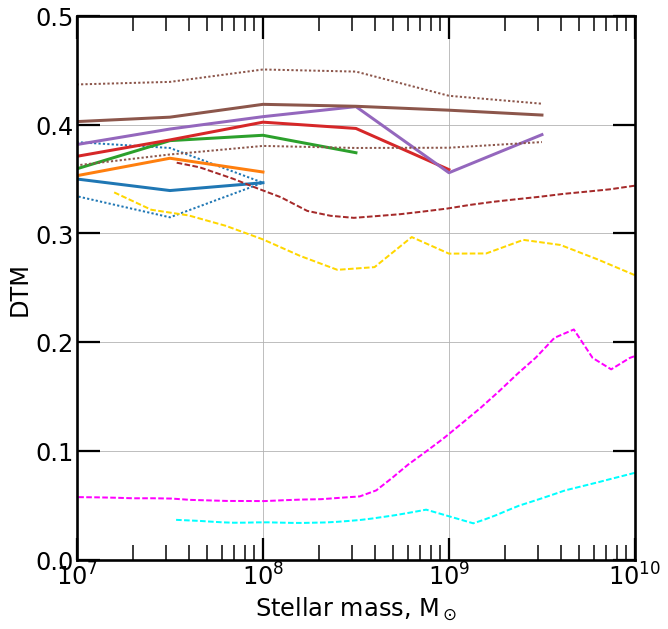}
    \caption{\emph{Top left:}The median relations between dust masses and stellar masses in galaxies. Full curves show the averages from our simulation at various redshifts, whereas dotted lines show the $\rm 16^{th}$ and $\rm 84^{th}$ percentile lines of the distribution at $\rm z=6$. Crosses show observational constraints gathered by \protect \cite{mancini_dust_2015,burgarella_observational_2020}, dashed lines show SAM predictions from \protect \cite{popping_dust_2017,vijayan_detailed_2019, dayal_alma_2022}. Finally, the grey area represents the results from the simulation of \protect \cite{graziani_assembly_2020}.
   \emph{Top right:} Median galactic dust to gas ratio (DTG) versus stellar mass. Full curves show the median from our simulation at various redshifts, whereas dotted lines show the 16$^{\rm th}$ and 84$^{\rm th}$ percentile lines of the distribution at $\rm z=6$. Dashed lines show SAM predictions from \protect \cite{popping_dust_2017,vijayan_detailed_2019, dayal_alma_2022}.
    \emph{Bottom Left:} Median galactic dust to metal ratio (DTM) versus stellar mass. Full curves show the medians from our simulation at various redshifts, whereas dotted lines show the $\rm 16^{ th}$ and $\rm 84^{ th}$ percentile lines of the distribution at z=6. Dashed lines show SAM predictions from \protect \cite{popping_dust_2017,vijayan_detailed_2019, dayal_alma_2022}.
    \emph{Bottom Right:} Median DTM for high stellar masses and in a linear scale. This panels highlights the evolution of the median DTM for high stellar mass galaxies that is obscured by the scaling of the top right panel.}
    \label{fig:gal_pties}
  \end{figure*}

Now we move to understand the galactic dust masses in relation to other galactic properties, such as stellar mass. The top left panel of Fig. \ref{fig:gal_pties} shows the median relation between dust mass and stellar mass in galaxies, with a collection of observational results and predictions from semi-analytical models and another simulation. The median dust mass increases with stellar mass for all stellar masses and at all redshifts. This is intuitive as our dust model includes the production of dust during the supernova events of stellar particles. Higher stellar mass galaxies in our simulation will tend to have experienced and to experience more supernova events and so produce more dust. One might expect that as time goes on, dust mass would increase on average at fixed stellar mass. However, this is not seen here. For the highest stellar mass haloes ($\rm M_\star > 10^8 M_\odot$) our median dust masses are a good match to the locus of observational points, as well as to the "high-cond" model of \cite{popping_dust_2017} at $\rm z= 6$ towards the end of Reionization. For the highest stellar mass galaxies, our predictions are also in quite good agreement with their "fiducial" model, but overshoot the results of \cite{vijayan_detailed_2019} and \cite{dayal_alma_2022} by almost a factor of 10.  There appears to be two regimes of dust accumulation, with a sudden increase of a factor $\sim 10^3$ in dust mass taking place around a stellar mass of  \SIrange{e5}{e6}{\simsun} (depending on the redshift). The existence of these two regimes is owed to the construction of our physical dust model. Whereas the high dust mass regime corresponds to galaxies in which the dust mass is limited by the maximum dust to metal ratio (set to 0.5 for every cell), the low dust mass regime corresponds to galaxies where accretion onto dust grains is inefficient and most dust mass originates directly from SNe ejecta without further growth (we confirmed this in a test simulation in which accretion onto dust grains was disabled). In fact, we can derive upper and lower limits (shown in dotted black lines in the top left panel of Fig. \ref{fig:gal_pties}) for the dust masses in our simulated galaxies by considering the total mass of metals deposited by SNe in the ISM\footnote{This approach is only valid for galaxies in which the mass of stellar particles younger than \SI{10}{\mega\year} is negligible when compared to the mass of older stellar particles, which is correct for massive galaxies.}, as follows:

\begin{equation}
    \left\{
\begin{array}{lll}
  \rm M_{metals} & \approx & \rm \eta y M_\star \, (assuming \, M_\star \approx M_\star[age>10Myr]) \,, \\
  \rm M_{dust, upper} & = & \rm max(DTM) M_{metals} \, , \\ 
  \\
  \rm M_{dust, lower} & = & \rm f_{cond} M_{metals}  \, (assuming \, no \, destruction) \, ,
\end{array}
\right.
\label{eq:dust_regimes}
\end{equation}
In the context of this toy model we assume all metals and dust are retained by the galaxies and there is no ejection into the IGM via galactic winds.

The resulting bounds $\rm M_{dust, upper} - \rm M_{dust, lower}$ neatly surround the \dustier{} dust masses, highlighting dust grain growth as the main cause of the regime change in dust production. Interestingly a similar shift in the dust mass to stellar mass relation is found by \cite{graziani_assembly_2020} in their simulations, and with a similar explanation (albeit at higher stellar masses). That the dust masses of galaxies can be so precisely determined (particularly by the upper limit for the high stellar mass galaxies) by these models explains the meagre evolution of the median dust masses at fixed stellar mass. It also suggests that dust destruction is not efficient in our galaxies, and that only a small fraction of the metals produced in our galaxies are ejected into the IGM.

Overall, and reassuringly, the dust masses of our most massive star forming galaxies seem quite realistic when compared to observations and other theoretical work.

To continue our investigation, we now turn to the median DTM of galaxies. To compute the DTM of a galaxy, we divide its total dust mass by its total metal mass (where the metal mass includes metals both in gas and in dust form). By this definition, $\rm DTM<1.0$. This also means that the values we shall be comparing are smoothed over the galaxies, even though individual cells in a galaxy can have very different local DTM. The top right panel of Fig. \ref{fig:gal_pties} shows the median galactic DTM as a function of galactic stellar mass. 

There are two striking aspects to these curves: Firstly, the median DTM essentially takes 2 main values, except between \SIrange{e5}{e6}{\simsun}, where it jumps abruptly from about  \num{e-3} to just under \num{0.4}. This reflects the two regimes seen in the dust mass - stellar mass relation described in the top left panel of Fig. \ref{fig:gal_pties} and happens at the same stellar mass: high stellar mass galaxies have high dust masses and high DTMs.

Secondly, in the high dust mass regime, {the median DTM only increases very little (by roughly 0.05) over the course of the simulation as shown by the bottom right panel.} As with the dust masses, there is very little scatter around the median DTM. The DTM of the two dust production regimes can be estimated in the same way we used previously, and by dividing the approximate dust masses given in Eq. \ref{eq:dust_regimes} by the approximate metal mass. Proceeding thus, we obtain an estimate for the DTM of each regime: \num{e-3} for the low dust regime where dust grain growth is inefficient and the expected DTM is the fraction of metals released by supernovae as dust ($\rm f_{cond}$); 0.5 for the high dust regime where the DTM of a galaxy is limited by the maximum DTM allowed in each cell (max(DTM))\footnote{Note that the limits for the galactic DTMs seem to bound the data much less tightly than the equivalent limits for the dust mass. This is because the computed DTMs are smoothed over each galaxy. i.e.: The highest DTM galaxies have DTMs just under 0.5 and contain many cells where DTM=max(DTM), however there remain cells with much lower DTMs.}. Again, our results are in a relatively close agreement with the predictions from the 'high-cond' model of \cite{popping_dust_2017} at $\rm z= 6$, and overshoot their 'fiducial' model and that of  \cite{vijayan_detailed_2019}. {In fact, for the highest stellar masses, we report median DTMs 0.1 higher than in the \cite{popping_dust_2017} 'high-cond' model, despite the excellent agreement in dust masses. This can be explained by lower metallicities in \dustier{} galaxies (0.5 dex lower for the highest stellar masses). Our results are similar to those of \cite{dayal_alma_2022}, who report lower dust masses at fixed stellar mass, thus implying higher metallicities in \dustier{}. } Note that the fiducial model of \cite{popping_dust_2017} and the model of \cite{vijayan_detailed_2019} both predict a jump in the average DTM as a function of stellar mass, but at higher stellar masses, with a slighter difference before and after the jumps. In both cases, the authors found that this jump in DTM is caused by an increase in dust grain growth, echoing our findings, but at lower stellar mass in our case.

The bottom left panel of Fig. \ref{fig:gal_pties} shows the median galactic dust to gas ratio (DTG) as a function of stellar mass. In our work, this is defined as the ratio between the dust mass and total gas mass of galaxies. The median DTG increases with stellar mass at all times, particularly sharply from \SIrange{e5}{e6}{\simsun} as does the median dust mass and median DTM. Again, this is driven by the increase in dust mass occurring as the accretion onto dust grains becomes more efficient in higher stellar mass galaxies. As with the other observables we have investigated, there is very little redshift evolution or scatter around the median. Whereas the median dust masses and DTM values agreed well with the predictions of the 'high-cond' model of \cite{popping_dust_2017} and \cite{dayal_alma_2022}, here we under-predict the median DTG when compared to the 'high-cond' \cite{popping_dust_2017} model, and end up with a slightly better agreement with the DTG from their 'fiducial' model. This discrepancy with respect to the \cite{popping_dust_2017} findings likely arises from the definition of DTG, and modelling of the hot and cold phases of the ISM. Indeed, \cite{popping_dust_2017} define DTG as the ratio between the dust mass, and the mass of neutral hydrogen and molecular hydrogen which is more faithful to its determinations in the lower redshift universe\citep[as in][]{ruyer_linking_2015}.


{
Over all, our agreement with observations and other modeling works on dust masses and their relation to stellar mass, gas mass, and metallicity is good enough for our purposes, and within a broad range explored by models and allowed by the (arguably limited) observational data at high redshifts. In the most massive star forming galaxies in the \dustier{} volume, we find dust masses within the upper limits given by observations, and comparable to some of the highest theoretical predictions in the literature. These galaxies are able to efficiently grow dust grains from the available gaseous metals, yielding high DTM values \citep[as also reported in ][]{popping_dust_2017, graziani_assembly_2020}. In practice, our model sets an upper limit for the DTM in every cell, allowing us rough control over the maximum expected galactic dust mass for a given stellar mass. We choose a high DTM limit of 0.5, aiming to remain compatible with observations, whilst allowing us to explore a scenario with very high dust masses, and thus providing an upper bound on the effects of dust on Reionization. 
For $\rm M_\star \leq 10^6 M_\odot$ galaxies, we report very low dust to stellar mass ratios and DTMs. In these galaxies, dust grain growth is inefficient and the dust masses are set by our choice of $\rm f_{cond}$, the dust mass fraction of supernovae metal ejecta. For these fainter galaxies, there are no high redshift constraints on dust masses or other related properties. 
In practice, our choice of $\rm f_{cond}$ is considerably lower than typically considered by theoretical works \citep[typically >0.01 e.g.][]{bianchi_dust_2007}. Our motivations for this choice are as follows: First, picking a significantly larger $\rm f_{cond}$ leads to a far gentler rise of the cosmic dust density when compared to semi-analytical modeling. Second, dust destruction processes are inefficient in \dustier{}'s galaxies because their ISM is not spatially  resolved. Setting a very low value of $\rm f_{cond}$ as we do can be thought of as a attempt to compensate for the inefficiency of dust destruction. 

}
\subsection{Reddening and extinction}
\label{sec:pred}
\paragraph*{Reddening of the UV continua of galaxies}


In the previous paragraphs, we validated our dust model using direct observational estimates of dust masses at high redshifts and models from the literature. Here we investigate how well the reddened star light properties of our simulated galaxies match existing observables, in particular the UV slope ($\beta$) and the UV extinction.

\begin{figure*}
  \includegraphics[trim={0cm 0cm 0.cm 0cm},clip,width=0.95\columnwidth]{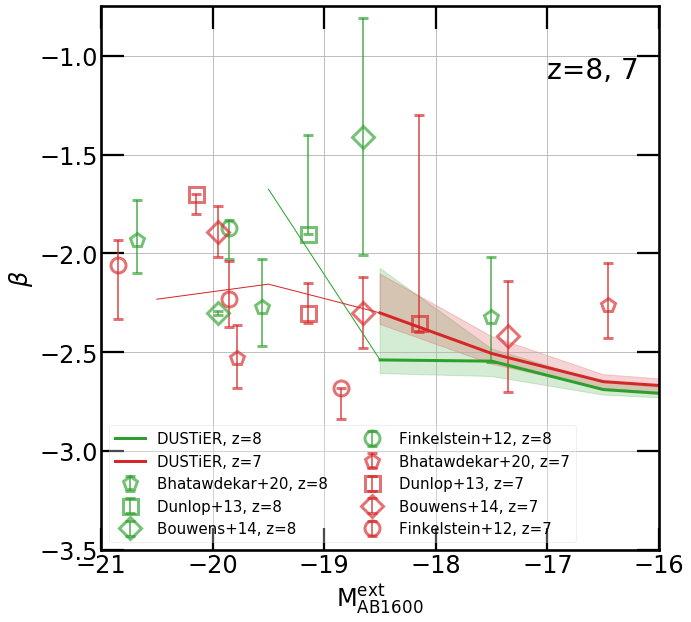}
  \includegraphics[trim={0cm 0cm 0.cm 0cm},clip,width=0.95\columnwidth]{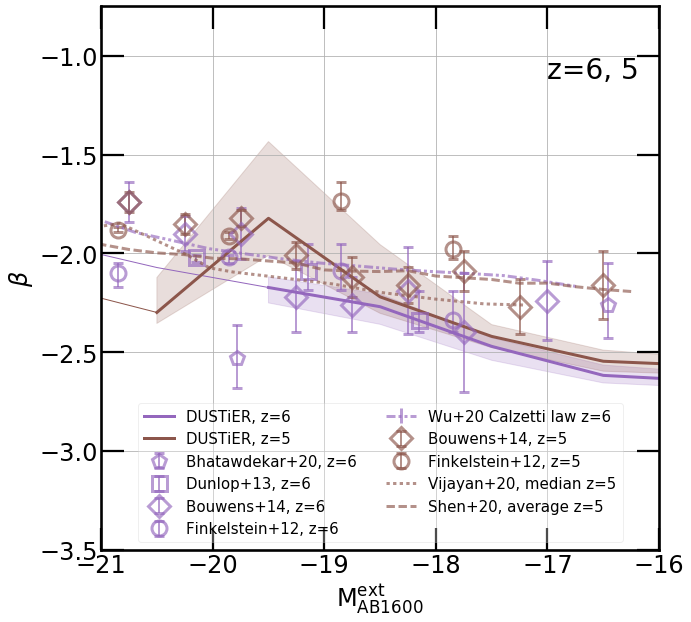}
    \caption{Coloured full lines show the evolution of the median relation between $\beta$ and \magyext \, in \dustier{} for \magyext \, bins with more than or exactly 5 galaxies (the thin, dashed continuation of these lines shows the medians of bins with fewer galaxies). These results use the LMC extinction curve from \protect \cite{draine_infrared_2001}. Coloured areas show the inter quartile range for each \magyext \, bin. Squares, circles, diamonds, and pentagons show constraints from observations \protect \citep{finkelstein_candels_2012,bouwens_census_2014,dunlop_uv_2013,bhatawdekar_uv_2020}. Models from \protect \cite{vijayan_first_2020,shen_high-redshift_2020,wu_photometric_2020} are shown in thick dotted, dot dashed, and dashed lines. To produce the points representing the \protect \cite{finkelstein_candels_2012,dunlop_uv_2013}, we re-binned their sample using the same \magyext \, bins as when binning the \dustier{} data, for consistency. The points and errorbars denote the median value and the error on the median estimated by bootstrapping the data (similar to the methods employed in the respective papers). Note that although \dustier{} \magyext \, stretch all the way to $\approx -10$, we only show the results for the brightest galaxies that have observational constraints.}
    \label{fig:betas}
  \end{figure*}

The two panels of Fig. \ref{fig:betas} show the median evolution of the UV slope ($\beta$) as a function of \magyext \, over time in \dustier{} (using the LMC values of dust attenuation coefficients ($\kappa_{d}$) from \cite{draine_infrared_2001}), compared with the evolution from the observations of \cite{finkelstein_candels_2012,bouwens_census_2014,dunlop_uv_2013,bhatawdekar_uv_2020}. At all times, we find that bright simulated galaxies are redder (shallower UV slope) than their fainter counterpart. {Roughly speaking the UV slope reddens by $\approx 0.6$ between \magyext{}$=-16$ and $-20.5$. For most magnitudes, the median UV slope ($\beta$) increases slightly (<0.1) with redshift. For some of the brightest galaxies ($\rm \approx -19.5$), the median $\beta$ varies by a significant margin between snapshots (>0.25). This can be attributed to both the modest box size of \dustier{}, and intrinsic variations in reddening at fixed magnitude. For instance, at z$=5$, the median slope increases from -2.25 to -1.8 then back again to -2.25 for \magyext$\in [-20.5,-18.5]$. It's probable that the "real" median UV slope lies somewhere between these values.}

Given the fairly large intrinsic scatter in our data at the bright end, as well as the large error bars given by observational constraints, our results seem to be in relatively good agreement with the former, especially for \magyext \, $ < -18$ objects. However the picture is unclear due to the aforementioned small sample of bright \dustier{} galaxies. This good agreement could be surprising as the dust masses of these galaxies are rather large. It may be that our chosen set of $\kappa_{d}$ are unrealistic for our high-redshift galaxies. However, because the high-redshift extinction law is unconstrained, we must content ourselves with choosing the one that allows us to best reproduce the constraints on the UV slopes (as done elsewhere e.g. \cite{vijayan_first_2020}). Our UV slopes for the $-16.5\geq$\magyext{}$\geq-17.5$ consistently lie below constraints. {This discrepancy may have to do with our lack of nebular emission modelling, as \citep[][showed that this process can redden $\beta$ slopes by roughly 0.1-0.3]{wilkins_photon_2016}. However, the current tension is mild (<0.2), and we prefer to focus on matching the constraints for brighter galaxies for which observational constraints are slightly more numerous (although displaying a large amount of dispersion).} Other models we compare our work with find gentler relations between \magyext \, and $\beta$ (akin to some linear function of \magyext). {This may be noteworthy, as theses studies do not attempt to directly use computed dust densities in their photometric computations, and adopt a more conservative approach based on metallicities.}

Overall, when accounting for the large scatter in the observations and simulation data, \dustier{} yields fairly realistic UV slopes ($\beta$), and plausible, albeit steep, median relations between $\beta$ and \magyext.  Choosing a different extinction curve (SMC over LMC for instance) has been investigated, but in our case the LMC appeared as a better choice as it reproduced the magnitude vs UV slope slightly better for the redshifts we simulate, as shown in App. \ref{app:kappa_pick}.

\begin{figure}
  \includegraphics[trim={0cm 0cm 0.cm 0cm},clip,width=0.95\columnwidth]{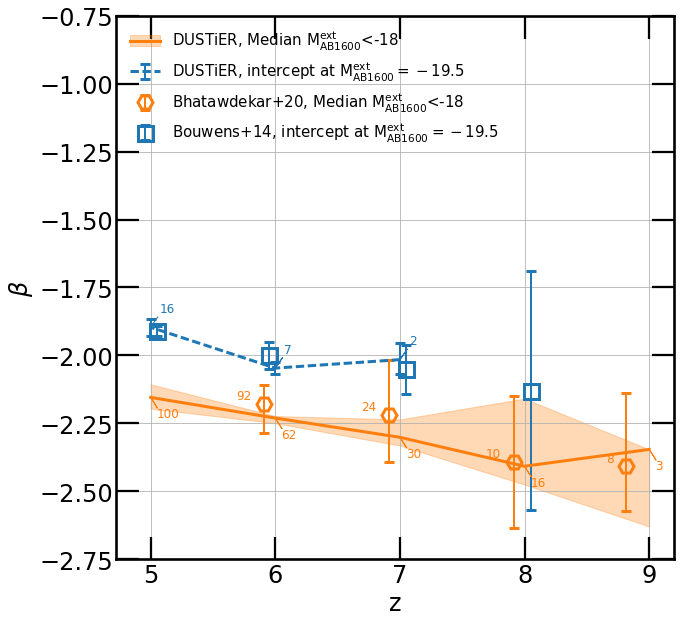}
    \caption{The full line shows the median UV slopes ($\beta$ ) of bright (\magy <-18) galaxies as a function of time in \dustier{}, the corresponding shaded area represents estimated error on the median obtained by bootstrapping our sample (similar to what is done in \protect \cite{bhatawdekar_evolution_2019}, and denoted by orange hexagons). The dashed line shows the intercept at \magyext$=-19.5$ of the fitted $\beta-$\magyext \, relation for several redshifts, comparable to the values reported by \protect \cite{bouwens_uv-continuum_2014} (green squares). The errorbars were also obtained by bootstrapping.  {Where possible, we have added the sample size of galaxies for each data point. We have omitted redshifts where the result depended on 2 or fewer galaxies.}}
    \label{fig:med_betas}
  \end{figure}

Let us turn to the temporal evolution of the UV slope of our simulated galaxies.
Fig. \ref{fig:med_betas} shows the median evolution of the $\beta$-\magyext \, relation with redshift. It shows the evolution of two separate metrics inspired by the literature : the bootstrapped median UV slope (as in \cite{bhatawdekar_uv_2020}), and the intercept of the fitted\footnote{linear least mean squares fit for galaxies with \magyext{}$<-17$} $\beta-$\magyext \, relation at \magyext$=-19.5$ (as in \cite{bouwens_uv-continuum_2014}). Our data suggests that the UV slopes of bright galaxies show a slight, evolution over time, increasing (reddening) between $\rm z=9$ and $\rm z=5$ {($\approx 0.45$ for the median <-18 curve), depending somewhat on the chosen metric}. More reddening in bright galaxies can be readily understood, since over time, as the number and mass of massive galaxies increases, the number of galaxies extincted down to \magyext$=-19.5$ is likely to increase. Thereby potentially increasing the fraction of galaxies at this magnitude that are heavily extincted and reddened.{The agreement is generally very good for both computed metrics, except at $\rm z=9$ for the median \magyext{}<-18 curve, where we lie at the bottom edge of the constraint from \cite{bhatawdekar_uv_2020}. However, these observational constraints rely on some of the brightest galaxies that are poorly represented in \dustier{} due to its volume. In fact, at some redshifts \dustier{} even has a smaller sample of comparable galaxies.}


Having demonstrated that our dust model produces {high, but plausible} dust masses in high redshift massive star forming galaxies, as well as reasonable\footnote{though admittedly predicting somewhat bluer \magyext{}$\geq-17.5$ galaxies} reddening of the UV continuum, we now turn to predictions regarding the effect of dust on the UVLF, and on the escape fraction of ionising light from galaxies.

\paragraph*{Extinction}

\begin{figure}
    \includegraphics[trim={0cm 0cm 0.cm 0cm},clip,width=\columnwidth]{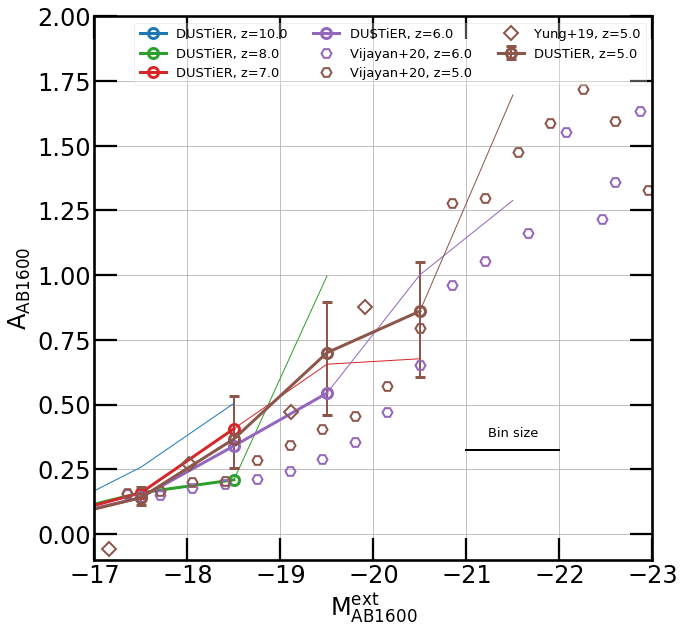}
    \caption{Median extinction at \SI{1600}{\angs} as a function of \magyext. Thick solid lines show the median value in bins with more than 5 galaxies, whereas thin lines show the median values in bins with less or exactly 5 galaxies. Error bars show the extent of the region between 16\% and 84\% values of extinction. For reference we show comparable results from \protect \cite{vijayan_first_2020} (circles) and from the SAM of \protect \cite{yung_semi_2019}.} 
    \label{fig:extinction}%
    \end{figure}

First, we compute the \SI{1600}{\angs} dust extinction \extinct \ of our simulated galaxies as \extinct $\rm= M^{ext}_{AB1600} - M^{int}_{AB1600}$, where \magyint \, is the intrinsic (i.e. with no reddening) absolute magnitude of a galaxy, and \magyext \, the magnitude accounting for extinction. The resulting median dust extinction is shown in Fig. \ref{fig:extinction}. The median \extinct \, increases substantially with decreasing \magyext \, at all redshifts, going from close to {0.1 at -17 to around 0.8 near -20.5} at $\rm z=5$. 
For every redshift, the most extincted galaxies are the brightest. The scatter around the median \extinct \, value also increases towards brighter galaxies. The error bars at \magyext=-20.5 show that the distribution of \extinct \, becomes very wide{, and stretching further below the median \extinct \, value than above it. This wide scatter is reminiscent of the strong variation in UV slope for a given magnitude. In part, this could be the sign of LoS variability. Though typical galaxies at \magyext=-20.5 have extinction values close to 0.8, there are a few galaxies for which the column density of dust along the simulated LoS is far smaller, giving \extinct \, values as much as 0.5 dex lower (equivalent to close to a factor $\approx0.6$ difference in observed luminosity at \SI{1600}{\angs})}. \cite{vijayan_first_2020} report a gentler slope of the relation between \extinct \, and \magyext{}, {which is unsurprising as their galaxies have lower dust masses. That being said,} considering the large scatter in the distribution of \dustier{} \extinct values, the discrepancy is mild for \magy $\rm \gtrsim -20$. Our median results are very consistent with those of \cite{yung_semi_2019} at $\rm z=5$ and for \magyext$\leq18$.

\paragraph*{Extinction and the UVLF}

\begin{figure*}
    \includegraphics[trim={0cm 0cm 0.cm 0cm},clip,width=\columnwidth]{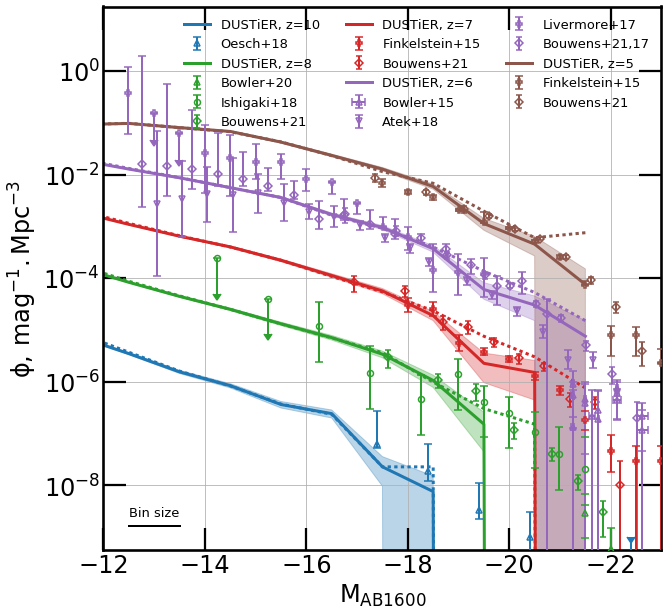}
    \includegraphics[trim={0cm 0cm 0.cm 0cm},clip,width=\columnwidth]{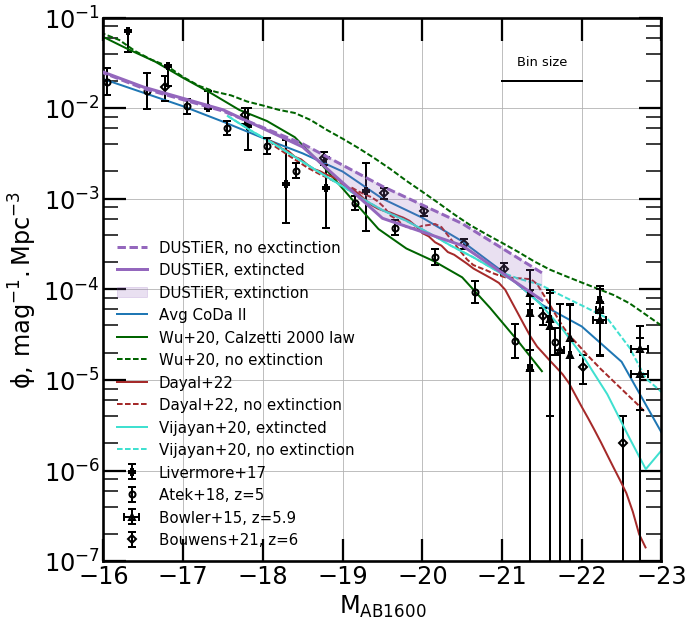}
    \caption{\emph{Left}: UVLF at various redshifts. Full lines show the UVLF when accounting for UV extinction, and dashed lines show the UVLF assuming no dust or extinction. The shaded areas show the poissonian error around the full lines. The small hollow diamonds with error bars were taken from \protect \cite{bouwens_uv_2015}. Note that for clarity we have shifted each successive set of curves downwards by one dex (so $\rm z=6$ points have their 'real' values, $\rm z=7$ data are 10x lower than in \dustier{}, $\rm z=8$ data are 100x lower than in \dustier{} and so on). \emph{Right}: extincted (full blue thick lines) and non-extincted (dashed blue thick lines) UVLFs in \dustier{} compared with observational constraints from \protect \cite{bouwens_uv_2015,bouwens_z_2017,livermore_directly_2017} (red and green markers), the \codaii \, UVLF (orange line). Similar results from the simulations of \protect \cite{wu_photometric_2020,vijayan_first_2020} are shown in brown and purple (again dashed lines for extincted UVLFs, and full lines for non-extincted UVLFs). We also show the results from the SAM Delphi \protect \cite{dayal_alma_2022}. Extincted UVLFs are a function of \magyext, whereas non extincted UVLFs are a function of \magyint.}
    \label{fig:UVLF}
  \end{figure*}

The left panel of Fig. \ref{fig:UVLF} shows the \dustier{} UVLF at various redshifts during Reionization between $\rm z=10$ and $\rm z=6$. At all times the UVLF takes the expected characteristic shape driven by hierarchical structure formation, with bright galaxies being rarer than faint ones. 
The dotted lines show the UVLF in \dustier{} when we consider no dust and no extinction, whereas the solid lines show the extincted \dustier{} UVLF. Strikingly, the UVLF is measurably affected by extinction even at very high redshift (Even at $\rm z=10$) for some of the brightest galaxies (\magy \, $<-19$), as indicated by the high redshift high median \extinct \, from Fig. \ref{fig:extinction}.
The abundance of the brightest galaxies (\magy{<-18.5}) in our simulation is strongly reduced when accounting for dust, and in some cases the corresponding magnitude bins are completely emptied (e.g. \magyext<-19 at $\rm z=8$).
In the bins where both the extincted and non-extincted UVLFs contain galaxies, for instance between -20<\magy<-19, the extincted UVLF can be modified by as much as $\lesssim$ $0.3$ dex, i.e. slightly more than a factor 2. Focusing on $\rm z=6$ in the right panel, we see that for fainter than \magy$>-18.5$ there is little to no difference between the two \dustier{} UVLFs. Between $-19$ and $-20$ the difference between the two curves due to extinction increases to $\sim 0.3$.
We now compare our two \dustier{} data-sets to observations taken from \cite{bouwens_new_2021,atek_extreme_2018,oesch_hduv_2018,finkelstein_evolution_2015,livermore_directly_2017,ishigaki_full-data_2018} at the same redshifts. Broadly, the left panel of Fig. \ref{fig:UVLF} shows that for galaxies fainter than \magy$\approx-18.5$, the match between the \dustier{} extincted and non-extincted UVLFs with observations is always good. 
At the bright end (\magy<-20.5), though, the non-extincted UVLF tends to overshoot the observations at all redshifts (where available), particularly for redshifts of 6 and below. {Thus, the extinction we compute improves the agreement of our UVLF with constraints for the very brightest galaxies of \dustier{}. However, \dustier{}'s volume does not contain many of these bright galaxies, even at $\rm z=5$. Therefore the predicted UVLF carry large uncertainties. With a larger volume we may find extinction to be too strong. Indeed, for fainter magnitudes (particularly for \magy=-19.5 at $\rm z=7$), extinction does appear too strong, and can push the extincted UVLF below observational constraints.}

The right panel zooms on the bright side of the UVLFs at $\rm z=6$. This more detailed view confirms that at the brightest magnitudes, the extincted UVLF is in better agreement with the observed data from \cite{bouwens_new_2021,livermore_directly_2017}. For reference we also show the UVLF from \codaii \, at this redshift, which does not account for extinction. The \codaii \, UVLF overshoots observations for the brightest galaxies (\magy < 21 from \cite{bouwens_new_2021}), and our results show that this could be resolved by dust extinction. In this panel, we also show similar results from the models of \cite{wu_photometric_2020,vijayan_first_2020}. We find \dustier{} sits at an intermediate position in terms of the degree with which extinction affects the UVLF: whereas the impact on the UVLF occurs for \magy<-18.5 in \dustier{}, in the simulations of \cite{wu_photometric_2020} it occurs as soon as \magy $\sim$ -17.5, and in the simulations of \cite{vijayan_first_2020} it occurs as late as -21. \cite{dayal_alma_2022} find a result close to that of \cite{vijayan_first_2020}, but with considerably fewer \magy $\rm<-21$ galaxies. The fact that \cite{vijayan_first_2020} observe that the UVLF is only extincted in brighter galaxies than in \dustier{} is intriguing since Fig. \ref{fig:extinction} showed that the median relation between extinction and magnitude was similar in both simulations. It could be that although the median \extinct \, is similar in both studies, \dustier{} contains a larger dispersion of extinction values than the model of \cite{vijayan_first_2020}. A small number of highly extincted galaxies could suffice to significantly affect the UVLF in the brightest \magy \, bins.

Overall, we find that the UVLF in \dustier{} is a good match to observational constraints at high redshift. This is owed, in part, to the extinction that we compute in post-processing, which improves the agreement for the brightest galaxies, and that has a dramatic effect even at high redshift. As we have shown in Sec. \ref{sec:dust_gals}, the massive galaxies in \dustier{} lie on the upper limit of dust masses compatible with observations. Thus, it is not surprising to find a high impact of extinction on UVLF. At the same time, one must consider the modest size of the simulation box, which could bias the result one way or another (massive galaxies could be under or over represented with respect to the average).

\paragraph*{Obscured SF}

Calibrated relations \citep[such as the one found in][]{madau_star_1998} can be used to infer the star formation rate density (SFRD) across time using constraints on the UVLF \citep[e.g.:][]{bouwens_census_2014,bouwens_uv_2015}. The canonical conversion that is employed is : $\rm L_{UV} \propto \frac{SFR}{M_\odot yr^{-1}} \, ergs\, s^{-1} \, Hz^{-1}$  \citep{madau_star_1998}. The value of $\rm L_{UV}$ must be corrected to take into account the extinction by dust. We set about studying the corrections made to account for dust in observations of the UVLF and the effect of dust predicted by \dustier{}, and the potential implications for the SFRD during Reionization.

\dustier{} gives us access to both the extincted and non-extincted UVLFs. We derive the fraction of obscured star formation $\rm f_{obs}$ (or the fraction of star formation missed because of the extinction of UV light) as follows for \dustier{} data :

\begin{equation}
  \rm f_{obs} = 1-\frac{L_{UV}}{L^{corr}_{UV}} \, ,
  \label{eq:SFR_Luv}
\end{equation}
where $\rm L_{UV}$ is the total integrated UV luminosity\footnote{computed as the integral of the UVLF weighted by luminosity} from galaxies at a given redshift, and $\rm L^{corr}_{UV}$ is the de-reddened, i.e. dust-corrected integrated UV luminosity. While difficult to obtain through observations, in \dustier{}, $\rm L^{corr}_{UV}$ is simply the integrated UV luminosity if we neglect the impact of dust, or the intrinsic UV luminosity $\rm L^{int}_{UV}$. Therefore, to be clear, we have $\rm L^{int}_{UV}$ = $\rm L^{corr}_{UV} \geq L_{UV}$.

  \begin{figure}
    \includegraphics[trim={0cm 0cm 0.cm 0cm},clip,width=\columnwidth]{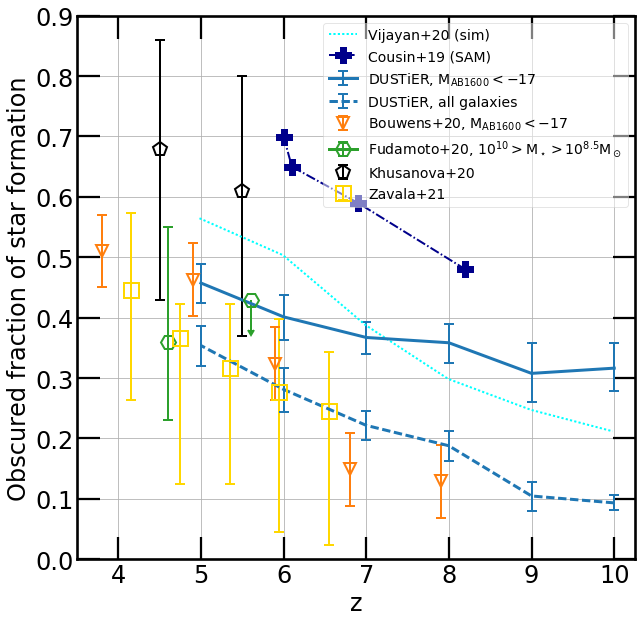}
    \caption{$\rm f_{obs}$ determined using Eq. \ref{eq:SFR_Luv}. Markers show $\rm f_{obs}$ results using observational data from \protect \cite{bouwens_alma_2020,fudamoto_alpine-alma_2020,zavala_evolution_2021,khusanova_alpine-alma_2020}. The solid blue line shows $\rm f_{obs}$ in \dustier{}. To allow for comparison with the results of \protect \cite{bouwens_alma_2020}, we only integrate the UVLF over halo magnitudes brighter than -17. The dashed blue line shows the \dustier{} result without a magnitude cut. For the \dustier{} curves, error bars show an error estimation produced by bootstrapping our procedure. We also include predictions from the SAM of \protect \cite{cousin_gs_2019} and the simulation of \protect \cite{vijayan_first_2020}.} 
    \label{fig:obscure}%
    \end{figure}

Fig. \ref{fig:obscure} shows $\rm f_{obs}$ in \dustier{} and for \cite{bouwens_alma_2020,fudamoto_alpine-alma_2020,zavala_evolution_2021,khusanova_alpine-alma_2020,cousin_gs_2019}. Just as reported by \cite{bouwens_alma_2020,zavala_evolution_2021}, the total fraction of star formation that is obscured by dust rises over time for both \dustier{} curves. For instance, for the \dustier{} \magy $<-17$ curve (that we'll call the "bright \dustier{} curve" from now on),  $\rm f_{obs}$ rises from just over 0.35 at $\rm z=8$ to around 0.45 at $\rm z=5$. This can be understood intuitively as the quantities we compute are luminosity-weighted and biased towards the most luminous, massive galaxies that have the largest dust masses and the most extinction. Over time more and more massive galaxies form. As these galaxies are less susceptible to star formation suppression during Reionization than low mass galaxies (which have only small dust masses), the total fraction of stellar mass in massive dusty galaxies can increase, and so can the fraction of obscured star formation. Similar trends are visible for most of the plotted constraints and results.

We also show the obscured fraction of star formation when no magnitude cut is done (the dashed blue \dustier{} curve), and the difference with the "bright \dustier{} curve" is dramatic. Indeed, as could have been expected, the \dustier{} bright curve represents galaxies with higher extinction, and thus higher $\rm f_{obs}$ values. The choice of sample also affects the redshift evolution of $\rm f_{obs}$. Whereas the full sample's (i.e. no mag cut) $\rm f_{obs}$ rapidly grows with decreasing redshift from  $\approx 0.1$ at $\rm z =9$ to $\approx 0.3$ at $\rm z=5$, the bright sample $\rm f_{obs}$ fluctuates around a values of $\approx 0.33$ for $\rm z\geq8$. This is because the amount of galactic dust and extinction in the bright sample has a lower bound that does not evolve much with redshift, which is not the case for the whole sample. The extent of the differences between the two \dustier{} curves illustrates the possible significant impact of selection effects and sample completeness on observational estimates of $\rm f_{obs}$.

At $\rm z=5$, the match between the highest constraints from observations and the \dustier{} bright sample is quite good. However, at higher redshifts the agreement deteriorates: for $\rm z \geq 6$ in the \dustier{} bright sample, $\rm f_{obs}$ is systematically higher than observations by a significant margin ($\rm \gtrsim 0.1$). This appears consistent with the rest of our results, which present high extinction and reddening.

At the same time, there could be issues with our comparison to observational constraints. Indeed, \cite{khusanova_alpine-alma_2020} discuss the potential biases towards unobscured galaxies caused by targeting fUV detected galaxies. In fact \cite{fudamoto_alpine-alma_2020} and \cite{khusanova_alpine-alma_2020} both rely on the same ALMA survey, except the latter attempts to account for obscuration in undetected faint (because of extinction) galaxies, hence their much higher constraints and significantly wider error bars. Conversely, due to the modest box size of \dustier{}, it does not contain some of the brightest, SFR and dust rich galaxies observed in \cite{bouwens_alma_2020,fudamoto_alpine-alma_2020,khusanova_alpine-alma_2020}. Therefore, since dust masses and extinction increase with halo and stellar mass in our simulation, a larger simulated volume with more massive galaxies could lead to even higher predictions of $\rm f_{obs}$. {The comparison with \cite{vijayan_first_2020} is interesting: for $\rm z\leq7$, they predict higher $\rm f_{obs}$, most likely due to the lack of very bright galaxies in \dustier{}: Conversely for $\rm z\geq7$, $\rm f_{obs}$ is higher in \dustier{} in which the range of galaxies that are extincted extends to fainter magnitudes.}
We find that the whole sample (i.e. no mag cut) \dustier{} $\rm f_{obs}$  is very close to results from \cite{fudamoto_alpine-alma_2020,bouwens_alma_2020,zavala_evolution_2021}, suggesting that either these observations underestimate the extinction in high redshift galaxies, or/and that massive galaxies in \dustier{} are too extincted {(likelier as they are have very high dust masses)}. 
{Overall, the fact that our results are in broad agreement with the available literature, considering the wide observational uncertainties, is positive. Broadly, \dustier{}'s predictions are plausible, but are in some tension with constraints, that favour less extinction, lower dust masses in massive galaxies, and moderately redder UV slopes for \magyext{}>-17.5 galaxies. To some extent this could be an issue with \dustier{}'s sample of bright galaxies, and could be alleviated in a larger volume with more of these objects. In light of this, we proceed to evaluate the impact of dust extinction on cosmic reionization, keeping in mind that it most likely constitutes an upper limit.}

\subsection{Implications for Reionization: Escape fractions through dust}

\begin{figure*}
    \includegraphics[trim={0cm 0.cm 0cm 0cm},clip,width=\columnwidth]{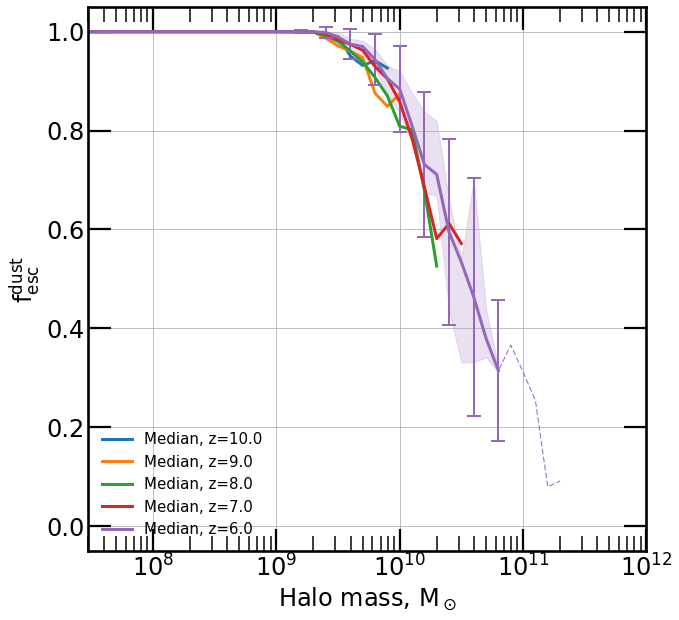}
    \includegraphics[trim={0cm 0cm 0cm 0.2cm},clip,width=\columnwidth]{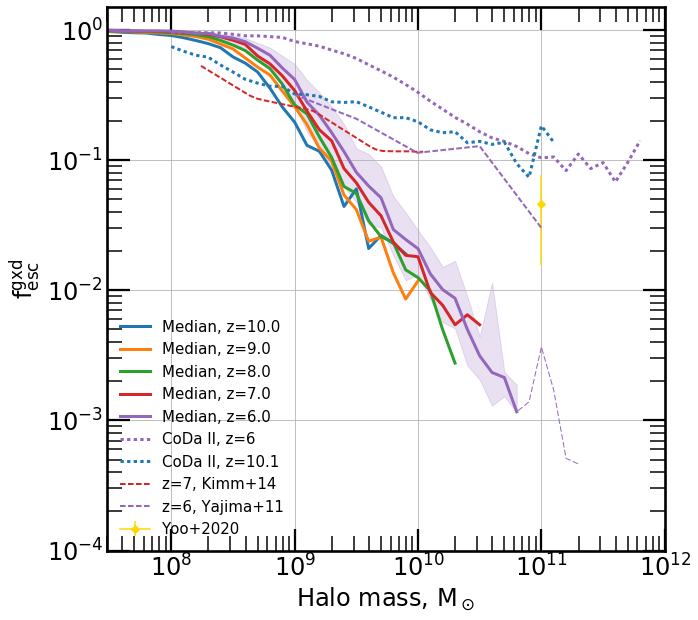}   
    \caption{\emph{Left} :  Median galactic escape fraction of ionising radiation \fescd as a function of mass, when only accounting for absorption by dust grains. Error-bars represent the standard deviation, whilst the shaded area represents the 16\% to 84\% regions of the distribution. Note the linear scale of the vertical axis. \emph{Right} : Median galactic escape fraction of ionising radiation \fesct as a function of mass, when accounting for absorption by both neutral hydrogen gas and dust grains. For reference we show similar results from \protect \cite{lewis_galactic_2020} as well as from \protect \cite{kimm_escape_2014,yajima_escape_2011,yoo_origin_2020}. Error-bars represent the standard deviation, whilst the shaded area represents the 16\% to 84\% regions of the distribution. Note the logarithmic scale of the vertical axis.}
    \label{fig:dust_fesc}
  \end{figure*}

In order to compute the loss of photons due to dust as they travel from star forming cells in the \rtwo \, of galaxies to the IGM, we use Eq. \ref{eq:dust_tau} to obtain the opacity due to dust along $\rm N_{LoS}$ for every star forming cell of every galaxy. For each galaxy, we then obtain the escape fraction of photons through dust by performing an angular average and an average over star forming cells weighted by their ionising luminosity \citep[inspired by the computation of the escape fractions as done in][]{lewis_galactic_2020}. 
The left panel of Fig. \ref{fig:dust_fesc} shows the median ionising escape fraction due to dust (or \fescd) as a function of halo mass for several redshifts in \dustier{}.  The most massive haloes have the lowest \fescd \, values. This is what we expect since the most massive haloes accrete the most gas and form the most stars, and thereby have the highest metal and dust masses. In haloes with masses lower than \SI{e10}{\simsun}, \fescd \, is very close to 1.0: dust has little or no effect (less than 10\%) on the contribution of ionising photons to Reionization in low and intermediate mass haloes in \dustier{}. However, \fescd \, decreases rapidly between \SI{e10}{\simsun} and $\approx$\SI{e11}{\simsun}, going from around 0.9, to close to 0.2 at and about 0.1 for the most massive haloes in the simulation.

We find very little evolution of \fescd with redshift, which may seem surprising at first. Indeed, one might imagine that as time goes on, dust accumulates in haloes of a fixed mass, thereby increasing the potential absorption of ionising photons due to dust. However in \dustier{}, the average \fescd \, remains relatively constant over time at fixed halo mass (when accounting for the low number statistics for most massive haloes at each redshift). Fig. \ref{fig:gal_pties} showed us that on average dust mass does not increase with time at fixed stellar mass. It could well be that the \fescd \, of individual haloes does indeed decrease over time, but that the decrease in \fescd \, corresponds with increases in halo mass, stellar mass, and dust mass.

We should not consider \fescd \, alone. The interesting quantity in Reionization simulations is the total escape fraction of ionising photons resulting from the absorption due to both neutral hydrogen gas and due to dust grains, i.e. the product \fesct = \fescg $\times$ \fescd, where \fescg is the escape fraction due to neutral Hydrogen. The right panel of Fig. \ref{fig:dust_fesc} shows the median value of \fesct as a function of halo mass and across time. We observe a similar overall trend to that found in \cite{lewis_galactic_2020} (shown here in dotted lines): a high \fesct \, plateau for low mass galaxies, that fades into a downwards slope with halo mass for high mass galaxies near $\lesssim$ \SI{e9}{\simsun} and onward. However, the aforementioned slope is noticeably steeper than reported in \codaii \,. Indeed, whereas haloes of \SI{e11}{\simsun} were found to have \fesc $\sim$ \SI{e-1} at $\rm z=6$ in \codaii \, (which did not feature dust), in \dustier{} we find values up to 100 times lower. Although it is tempting to ascribe this dramatic difference wholly to the new dust absorption modelling, this is not the case as the measured values of \fescd \, are not low enough to explain the difference on their own. In fact the much lower \fesct \, values are driven by stronger absorption by neutral hydrogen in galaxies than in \codaii \,, as already shown in \cite{ocvirk_lyman-alpha_2021}, due to the new calibration of the star formation sub-grid model. 

We showed in \cite{lewis_galactic_2020} that the main galactic drivers of cosmic Reionization in \codaii \, reside in dark matter haloes between \SI{6e8}{\simsun} and \SI{3e10}{\simsun}. In \dustier{}, such galaxies have \fescd \, values close to one throughout Reionization, meaning that dust probably does not strongly affect the main drivers of Reionization. Moreover, the \dustier{} setup uses the new star formation calibration of \cite{ocvirk_lyman-alpha_2021}, which results in lower escape fractions for massive haloes than in \codaii \,. We may therefore expect the mass range of the main drivers of reionization to shift to even lower masses than in \cite{lewis_galactic_2020}, suggesting an even smaller impact of dust on the photon budget of reionization.

    \begin{figure}
    \includegraphics[trim={0cm 0cm 0.cm 0cm},clip,width=\columnwidth]{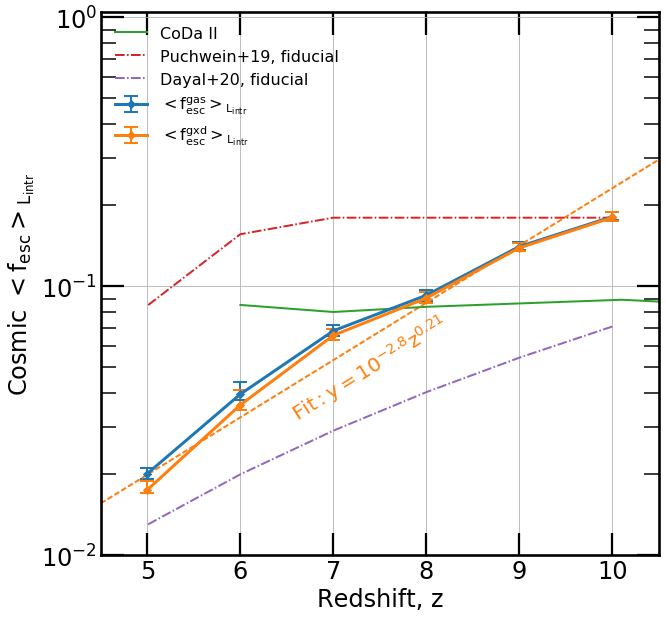}
    \caption{Average escape fraction: $\rm L_{intr}$ weighted average escape fraction of ionising photons due to absorption (. The full curves represent the median of the distribution of $\rm L_{intr}$ weighted average escape fractions, obtained by bootstrapping our sample of galaxies, whereas the error bars represent the 16\% to 84\% regions of the same distribution.}
    \label{fig:avg_z_dust_fesc}
  \end{figure}
  
As another quantity of interest, we now focus on the average "global" escape fraction. When considering cells so large that describing the detail of a galaxy population is not relevant or not useful, semi-analytical models of the Epoch of Reionization may often assume a constant global escape fraction, which, applied to the whole population of star-forming haloes, yields the cosmic emissivity.
  
Simulations such as \dustier{} are valuable as they are able to provide such a number. In this spirit, we define the cosmic average escape fraction as the average escape fraction of haloes weighted by their intrinsic ionising photon production $\rm L_{intr}$, i.e. 
\begin{equation}
    \rm <f_{esc}^{gxd}>_{L_{intr}} = \sum_i f_{esc,i}^{gxd} \, L_{intr,i} / \sum_i L_{intr,i} \, ,
\end{equation}
  
  where the index i runs over the population of haloes. By replacing \fesct by \fescg we also obtain the cosmic average escape fraction due to neutral hydrogen only $\rm <f_{esc}^{gas}>_{L_{intr}}$ (i.e. leaving out dust).

  We show the evolution of these cosmic escape fractions as a function of redshift in Fig. \ref{fig:avg_z_dust_fesc}. They decrease over time from close to 0.2 at $\rm z=10$ to just under 0.02 at $\rm z=5$. This decrease is imputable to both the build up of the number of massive galaxies with low \fesct \, values, and to the rise of star formation suppression in low mass galaxies: since the average is weighted by the intrinsic photon production of galaxies, quasi-proportional to the star formation rate, the average is biased increasingly strongly with time, against suppressed galaxies, and towards the most massive and luminous objects, which are the most opaque, as shown by \cite{ocvirk_lyman-alpha_2021} using a quasi identical model (but without dust). This decreasing trend with decreasing redshift is reminiscent of various models that have been suggested in the literature, such as by \cite{puchwein_consistent_2019,dayal_reionization_2020}, although in the latter, the decrease is driven by different physical processes.
  
  The cosmic average escape fraction including dust (orange solid line) is only slightly smaller than its neutral-hydrogen-only counterpart (blue solid line), although the difference increases towards low redshifts. At $\rm z=10$, dust has almost no effect on the fraction of escaping light. But by $\rm z=5$, accounting for dust reduces the total escaping photon fraction from $\sim 2\%$ to $\sim 1.8\%$. Crucially for our study of Reionization, this shows that the effect of dust on the total fraction of escaping ionising photons is very small. We fit a power law function to the cosmic escape fraction, finding that $<$\fesct$\mathrm{>_{L_{intr}}}=10^{-2.7}z^{0.21}$.
  
  Finally, we also show the cosmic average escape fraction computed for the \codaii \, simulation, which shows no evolution over the redshift range considered. Again, this is not due to our new dust implementation, but to the new calibration of the sub-grid star formation model, which makes high mass haloes much more opaque in \dustier{} than in \codaii.

  {We caution that the dust escape fractions derived in this section via post-processing, rely on an LMC extinction curve which we foudn to give a more realistic $\rm \beta -M_{1600,AB}$ relation. This is inconsistent with the extinction law from the SMC that was assumed during run time in the radiative transfer scheme of RAMSES-CUDATON. The swap from SMC to LMC extinction laws occurred after the simulation run, as iteration on our results led us to recognise that the LMC law gave us better agreements with observed reddening and extinction. Ideally, a new simulation would have been run using a LMC law at runtime and in post-processing/analysis, but our allocation eventually ran out. Also, this would have been very computationally expensive for relatively small gains. Though regrettable, we highlight that overall the impact of dust on Reionization in \dustier{} is small, we find it therefore preferable to introduce this slight inconsistency in order to present reasonable reddening, extinction properties for our galaxies. In App. \ref{app:kappa_pick} we explore the differences in our results when adopting either SMC or LMC extinction laws.}

\section{Conclusions}
\label{sec:concl}
We have coupled a new physically motivated dust model (see Dubois et al. in prep) to the RHD cosmological simulation code RAMSES-CUDATON, and performed the first cosmological simulations where dust production is coupled to both hydrodynamics and star formation, as well as the radiative transfer of ionising photons through hydrogen gas and dust. After calibrating the dust model, using other models found in the literature, the dust-related properties of our simulated galaxies are compatible with available high-redshift observations of dust masses.

Overall, we find that at fixed stellar mass, the dust properties of galaxies do not evolve significantly with time. In our model, there are two dust production channels: condensation in the ejecta of supernovae, and the growth onto existing dust grains in metal rich gas. We find that in galaxies with stellar masses lower than $\approx$ \SI{2e6}{\simsun}, accretion is inefficient, and the total mass of dust that condensates closely fits the dust masses of galaxies. Higher stellar mass galaxies are sufficiently enriched that dust grain growth becomes efficient, and the dust mass to stellar mass ratios of such galaxies are far higher.

Using a LMC extinction law, we compute the UV slope of our simulated galaxies and find a reasonable agreement with the observed $\beta$-\magyext \, relation at high redshift, {especially for the bright \dustier{} galaxies which are the most well constrained by observations.}

We study extinction at \SI{1600}{\angs} and its potential implications for UV observations of galaxies during Reionization. We find that dust produces measurable extinction of the UVLF as early as $\rm z=10$, and for galaxies brighter than -18.5. 

We also find an evolution of the UV slope with redshift compatible with observations. 

We estimate the the impact of dust in our simulation on UV based determinations of the cosmic star formation rate, and find that the fraction of obscured star formation increases over time, reaching 35\% to 45\% (depending on the chosen magnitude limit) in good agreement with the various observational constraints and other simulated work towards the end of Reionization.

Finally we address the influence of dust on the Reionization process itself. We show that dust can have a significant impact on the escape fraction of ionising photons of our galaxies above \SI{e10}{\simsun}, reducing the escaping ionising luminosity by a factor of $\approx$ 10\% that increases to $\approx$ 90\% by \SI{e11}{\simsun}. However, the absorption due to neutral Hydrogen in our galaxies is still the dominating contributor to low escape fractions in high mass galaxies, and we show that dust has a very moderate effect on the total fraction of escaping ionising photons, even at $\rm z=6$. This suggests that the presence of dust already during the Epoch of Reionization may not have a very significant effect on the timing or the topology of Reionization.

However, because of the modest box size of \dustier{}, we cannot relate the predictions of our dust model to the highest mass observational constraints available. At the same time, this means we cannot comment on the extinction of the brightest galaxies, nor can we investigate the \fescd \, in the brightest galaxies. That being said, \cite{lewis_galactic_2020} showed that the  >\SI{e11}{\simsun} galaxies were not the main drivers of Reionization in \codaii \,. With the added dust extinction we have measured here, they are even less likely to be important drivers of cosmic reionization, and therefore small-ish simulations such as \dustier{} may still be reasonable descriptions of cosmic reionization because the contribution due to the missing largest galaxies remains fairly small.
The new \codaiii \, simulation, which is currently in the early stages of analysis, will allow for a more in-depth investigation of these aspects, thanks to a significant step up in box size compared to \dustier{} (64 times larger in volume). In particular the larger box size will allow us to study larger and brighter galaxies with higher dust masses as well as bolster the number of high mass galaxies, allowing us to better compare the predictions of our model with other simulations, SAMs, and critically, observations. This will also allow us to perform a more detailed study of the effects of dust on the drivers of Reionization.

\section*{Acknowledgements}
This work was granted access to the HPC resources of IDRIS under the allocations 2019-A0090411049 and 2020-GC-JZ-CT4 made by GENCI.
This study is in part supported by the DFG via the Heidelberg Cluster of Excellence STRUCTURES in the framework of Germany’s Excellence Strategy (grant EXC-2181/1 - 390900948).
This work used v2.2.1 of the Binary
Population and Spectral Synthesis (BPASS) models as last presented
by \cite{eldridge_population_2020}. PO thanks H. Katz for suggesting the use of BPASS nebular emission line models to quantify their impact on galaxy photometric properties.
We also made use of Python, and the following packages : matplotlib \citep{Hunter:2007}, numpy \citep{van2011numpy}, scipy \citep{Virtanen_2020}, healpix \citep{gorski_HEALPix_2005}.

\section*{Data availability}
The data underlying this article will be shared on reasonable request
to the corresponding author.




\bibliographystyle{mnras}
\bibliography{calib_article.bib} 

\begin{thebibliography}{}
\makeatletter
\relax
\def\mn@urlcharsother{\let\do\@makeother \do\$\do\&\do\#\do\^\do\_\do\%\do\~}
\def\mn@doi{\begingroup\mn@urlcharsother \@ifnextchar [ {\mn@doi@}
  {\mn@doi@[]}}
\def\mn@doi@[#1]#2{\def\@tempa{#1}\ifx\@tempa\@empty \href
  {http://dx.doi.org/#2} {doi:#2}\else \href {http://dx.doi.org/#2} {#1}\fi
  \endgroup}
\def\mn@eprint#1#2{\mn@eprint@#1:#2::\@nil}
\def\mn@eprint@arXiv#1{\href {http://arxiv.org/abs/#1} {{\tt arXiv:#1}}}
\def\mn@eprint@dblp#1{\href {http://dblp.uni-trier.de/rec/bibtex/#1.xml}
  {dblp:#1}}
\def\mn@eprint@#1:#2:#3:#4\@nil{\def\@tempa {#1}\def\@tempb {#2}\def\@tempc
  {#3}\ifx \@tempc \@empty \let \@tempc \@tempb \let \@tempb \@tempa \fi \ifx
  \@tempb \@empty \def\@tempb {arXiv}\fi \@ifundefined
  {mn@eprint@\@tempb}{\@tempb:\@tempc}{\expandafter \expandafter \csname
  mn@eprint@\@tempb\endcsname \expandafter{\@tempc}}}

\bibitem[\protect\citeauthoryear{Atek, Richard, Kneib  \& Schaerer}{Atek
  et~al.}{2018}]{atek_extreme_2018}
Atek H.,  Richard J.,  Kneib J.-P.,   Schaerer D.,  2018, \mn@doi [Mon Not R
  Astron Soc] {10.1093/mnras/sty1820}, 479, 5184

\bibitem[\protect\citeauthoryear{Aubert \& Teyssier}{Aubert \&
  Teyssier}{2008}]{aubert_radiative_2008}
Aubert D.,  Teyssier R.,  2008, \mn@doi [Mon Not R Astron Soc]
  {10.1111/j.1365-2966.2008.13223.x}, 387, 295

\bibitem[\protect\citeauthoryear{Aubert \& Teyssier}{Aubert \&
  Teyssier}{2010}]{aubert_reionization_2010}
Aubert D.,  Teyssier R.,  2010, \mn@doi [The Astrophysical Journal]
  {10.1088/0004-637X/724/1/244}, 724, 244

\bibitem[\protect\citeauthoryear{Aubert et~al.,}{Aubert
  et~al.}{2018}]{aubert_inhomogeneous_2018}
Aubert D.,  et~al., 2018, \mn@doi [The Astrophysical Journal Letters]
  {10.3847/2041-8213/aab14d}, 856, L22

\bibitem[\protect\citeauthoryear{Bhatawdekar \& Conselice}{Bhatawdekar \&
  Conselice}{2020}]{bhatawdekar_uv_2020}
Bhatawdekar R.,  Conselice C.~J.,  2020, arXiv e-prints, 2006, arXiv:2006.00013

\bibitem[\protect\citeauthoryear{Bhatawdekar, Conselice, Margalef-Bentabol  \&
  Duncan}{Bhatawdekar et~al.}{2019}]{bhatawdekar_evolution_2019}
Bhatawdekar R.,  Conselice C.~J.,  Margalef-Bentabol B.,   Duncan K.,  2019,
  \mn@doi [Monthly Notices of the Royal Astronomical Society]
  {10.1093/mnras/stz866}, 486, 3805

\bibitem[\protect\citeauthoryear{{Bianchi} \& {Schneider}}{{Bianchi} \&
  {Schneider}}{2007}]{bianchi_dust_2007}
{Bianchi} S.,  {Schneider} R.,  2007, \mn@doi [\mnras]
  {10.1111/j.1365-2966.2007.11829.x}, \href
  {https://ui.adsabs.harvard.edu/abs/2007MNRAS.378..973B} {378, 973}

\bibitem[\protect\citeauthoryear{Bleuler, Teyssier, Carassou  \&
  Martizzi}{Bleuler et~al.}{2015}]{bleuler_phew_2015}
Bleuler A.,  Teyssier R.,  Carassou S.,   Martizzi D.,  2015, \mn@doi
  [Computational Astrophysics and Cosmology] {10.1186/s40668-015-0009-7}, 2, 5

\bibitem[\protect\citeauthoryear{Bouwens et~al.,}{Bouwens
  et~al.}{2014a}]{bouwens_uv-continuum_2014}
Bouwens R.~J.,  et~al., 2014a, \mn@doi [The Astrophysical Journal]
  {10.1088/0004-637X/793/2/115}, 793, 115

\bibitem[\protect\citeauthoryear{Bouwens et~al.,}{Bouwens
  et~al.}{2014b}]{bouwens_census_2014}
Bouwens R.~J.,  et~al., 2014b, \mn@doi [The Astrophysical Journal]
  {10.1088/0004-637X/795/2/126}, 795, 126

\bibitem[\protect\citeauthoryear{Bouwens et~al.,}{Bouwens
  et~al.}{2015}]{bouwens_uv_2015}
Bouwens R.~J.,  et~al., 2015, \mn@doi [ApJ] {10.1088/0004-637X/803/1/34}, 803,
  34

\bibitem[\protect\citeauthoryear{Bouwens, Oesch, Illingworth, Ellis  \&
  Stefanon}{Bouwens et~al.}{2017}]{bouwens_z_2017}
Bouwens R.~J.,  Oesch P.~A.,  Illingworth G.~D.,  Ellis R.~S.,   Stefanon M.,
  2017, \mn@doi [The Astrophysical Journal] {10.3847/1538-4357/aa70a4}, 843,
  129

\bibitem[\protect\citeauthoryear{Bouwens et~al.,}{Bouwens
  et~al.}{2020}]{bouwens_alma_2020}
Bouwens R.,  et~al., 2020, \mn@doi [The Astrophysical Journal]
  {10.3847/1538-4357/abb830}, 902, 112

\bibitem[\protect\citeauthoryear{Bouwens et~al.,}{Bouwens
  et~al.}{2021}]{bouwens_new_2021}
Bouwens R.~J.,  et~al., 2021, arXiv e-prints, 2102, arXiv:2102.07775

\bibitem[\protect\citeauthoryear{Burgarella, Nanni, Hirashita, Theule, Inoue
  \& Takeuchi}{Burgarella et~al.}{2020}]{burgarella_observational_2020}
Burgarella D.,  Nanni A.,  Hirashita H.,  Theule P.,  Inoue A.~K.,   Takeuchi
  T.~T.,  2020, arXiv, p. arXiv:2002.01858

\bibitem[\protect\citeauthoryear{Béthermin et~al.,}{Béthermin
  et~al.}{2015}]{bethermin_evolution_2015}
Béthermin M.,  et~al., 2015, \mn@doi [A\&A] {10.1051/0004-6361/201425031},
  573, A113

\bibitem[\protect\citeauthoryear{Calzetti, Kinney  \&
  Storchi-Bergmann}{Calzetti et~al.}{1994}]{calzetti_dust_1994}
Calzetti D.,  Kinney A.~L.,   Storchi-Bergmann T.,  1994, \mn@doi [The
  Astrophysical Journal] {10.1086/174346}, 429, 582

\bibitem[\protect\citeauthoryear{Cousin, Buat, Lagache  \& Bethermin}{Cousin
  et~al.}{2019}]{cousin_gs_2019}
Cousin M.,  Buat V.,  Lagache G.,   Bethermin M.,  2019, \mn@doi [Astronomy and
  Astrophysics] {10.1051/0004-6361/201834674}, 627, A132

\bibitem[\protect\citeauthoryear{Dawoodbhoy et~al.,}{Dawoodbhoy
  et~al.}{2018}]{dawoodbhoy_suppression_2018}
Dawoodbhoy T.,  et~al., 2018, \mn@doi [Monthly Notices of the Royal
  Astronomical Society] {10.1093/mnras/sty1945}, 480, 1740

\bibitem[\protect\citeauthoryear{Dayal \& Ferrara}{Dayal \&
  Ferrara}{2018}]{dayal_early_2018}
Dayal P.,  Ferrara A.,  2018, \mn@doi [PhR] {10.1016/j.physrep.2018.10.002},
  780, 1

\bibitem[\protect\citeauthoryear{Dayal et~al.,}{Dayal
  et~al.}{2020}]{dayal_reionization_2020}
Dayal P.,  et~al., 2020, arXiv, p. arXiv:2001.06021

\bibitem[\protect\citeauthoryear{{Dayal} et~al.,}{{Dayal}
  et~al.}{2022}]{dayal_alma_2022}
{Dayal} P.,  et~al., 2022, \mn@doi [\mnras] {10.1093/mnras/stac537}, \href
  {https://ui.adsabs.harvard.edu/abs/2022MNRAS.512..989D} {512, 989}

\bibitem[\protect\citeauthoryear{Deparis, Aubert, Ocvirk, Chardin  \&
  Lewis}{Deparis et~al.}{2019}]{deparis_impact_2019}
Deparis N.,  Aubert D.,  Ocvirk P.,  Chardin J.,   Lewis J.,  2019, \mn@doi
  [Astronomy and Astrophysics] {10.1051/0004-6361/201832889}, 622, A142

\bibitem[\protect\citeauthoryear{Draine \& Li}{Draine \&
  Li}{2001}]{draine_infrared_2001}
Draine B.~T.,  Li A.,  2001, \mn@doi [The Astrophysical Journal]
  {10.1086/320227}, 551, 807

\bibitem[\protect\citeauthoryear{Draine \& Salpeter}{Draine \&
  Salpeter}{1979}]{draine_destruction_1979}
Draine B.~T.,  Salpeter E.~E.,  1979, \mn@doi [The Astrophysical Journal]
  {10.1086/157206}, 231, 438

\bibitem[\protect\citeauthoryear{Dubois \& Teyssier}{Dubois \&
  Teyssier}{2008}]{dubois_supernova_2008}
Dubois Y.,  Teyssier R.,  2008. p.~388, \url
  {http://cdsads.u-strasbg.fr/abs/2008ASPC..390..388D}

\bibitem[\protect\citeauthoryear{Dunlop et~al.,}{Dunlop
  et~al.}{2013}]{dunlop_uv_2013}
Dunlop J.~S.,  et~al., 2013, \mn@doi [Monthly Notices of the Royal Astronomical
  Society] {10.1093/mnras/stt702}, 432, 3520

\bibitem[\protect\citeauthoryear{Dwek}{Dwek}{1998}]{dwek_evolution_1998}
Dwek E.,  1998, \mn@doi [The Astrophysical Journal] {10.1086/305829}, 501, 643

\bibitem[\protect\citeauthoryear{Eldridge \& Stanway}{Eldridge \&
  Stanway}{2020}]{eldridge_population_2020}
Eldridge J.~J.,  Stanway E.~R.,  2020, arXiv e-prints, 2005, arXiv:2005.11883

\bibitem[\protect\citeauthoryear{Erb, Shapley, Pettini, Steidel, Reddy  \&
  Adelberger}{Erb et~al.}{2006}]{erb_mass-metallicity_2006}
Erb D.~K.,  Shapley A.~E.,  Pettini M.,  Steidel C.~C.,  Reddy N.~A.,
  Adelberger K.~L.,  2006, \mn@doi [The Astrophysical Journal]
  {10.1086/503623}, 644, 813

\bibitem[\protect\citeauthoryear{Faisst et~al.,}{Faisst
  et~al.}{2016}]{faisst_rest-uv_2016}
Faisst A.~L.,  et~al., 2016, \mn@doi [The Astrophysical Journal]
  {10.3847/0004-637X/822/1/29}, 822, 29

\bibitem[\protect\citeauthoryear{{Ferland}, {Korista}, {Verner}, {Ferguson},
  {Kingdon}  \& {Verner}}{{Ferland} et~al.}{1998}]{ferland_cloudy_1998}
{Ferland} G.~J.,  {Korista} K.~T.,  {Verner} D.~A.,  {Ferguson} J.~W.,
  {Kingdon} J.~B.,   {Verner} E.~M.,  1998, \mn@doi [\pasp] {10.1086/316190},
  \href {https://ui.adsabs.harvard.edu/abs/1998PASP..110..761F} {110, 761}

\bibitem[\protect\citeauthoryear{Finkelstein et~al.,}{Finkelstein
  et~al.}{2012}]{finkelstein_candels_2012}
Finkelstein S.~L.,  et~al., 2012, \mn@doi [The Astrophysical Journal]
  {10.1088/0004-637X/756/2/164}, 756, 164

\bibitem[\protect\citeauthoryear{Finkelstein et~al.,}{Finkelstein
  et~al.}{2015}]{finkelstein_evolution_2015}
Finkelstein S.~L.,  et~al., 2015, \mn@doi [The Astrophysical Journal]
  {10.1088/0004-637X/810/1/71}, 810, 71

\bibitem[\protect\citeauthoryear{Fromang, Hennebelle  \& Teyssier}{Fromang
  et~al.}{2006}]{fromang_high_2006}
Fromang S.,  Hennebelle P.,   Teyssier R.,  2006, \mn@doi [Astronomy and
  Astrophysics] {10.1051/0004-6361:20065371}, 457, 371

\bibitem[\protect\citeauthoryear{Fudamoto et~al.,}{Fudamoto
  et~al.}{2020}]{fudamoto_alpine-alma_2020}
Fudamoto Y.,  et~al., 2020, \mn@doi [Astronomy and Astrophysics]
  {10.1051/0004-6361/202038163}, 643, A4

\bibitem[\protect\citeauthoryear{Gnedin}{Gnedin}{2016}]{gnedin_proper_2016}
Gnedin N.~Y.,  2016, \mn@doi [The Astrophysical Journal]
  {10.3847/1538-4357/833/1/66}, 833, 66

\bibitem[\protect\citeauthoryear{{G{\'o}rski}, {Hivon}, {Banday}, {Wand elt},
  {Hansen}, {Reinecke}  \& {Bartelmann}}{{G{\'o}rski}
  et~al.}{2005}]{gorski_HEALPix_2005}
{G{\'o}rski} K.~M.,  {Hivon} E.,  {Banday} A.~J.,  {Wand elt} B.~D.,  {Hansen}
  F.~K.,  {Reinecke} M.,   {Bartelmann} M.,  2005, \mn@doi [\apj]
  {10.1086/427976}, \href
  {https://ui.adsabs.harvard.edu/abs/2005ApJ...622..759G} {622, 759}

\bibitem[\protect\citeauthoryear{Graziani, Schneider, Ginolfi, Hunt, Maio,
  Glatzle  \& Ciardi}{Graziani et~al.}{2020}]{graziani_assembly_2020}
Graziani L.,  Schneider R.,  Ginolfi M.,  Hunt L.~K.,  Maio U.,  Glatzle M.,
  Ciardi B.,  2020, \mn@doi [Monthly Notices of the Royal Astronomical Society]
  {10.1093/mnras/staa796}, 494, 1071

\bibitem[\protect\citeauthoryear{Gronke et~al.,}{Gronke
  et~al.}{2020}]{gronke_lyman-alpha_2020}
Gronke M.,  et~al., 2020, arXiv e-prints, 2004, arXiv:2004.14496

\bibitem[\protect\citeauthoryear{He, Ricotti  \& Geen}{He
  et~al.}{2019}]{He2019}
He C.-C.,  Ricotti M.,   Geen S.,  2019, \mn@doi [Monthly Notices of the Royal
  Astronomical Society] {10.1093/mnras/stz2239}, 489, 1880

\bibitem[\protect\citeauthoryear{He, Ricotti  \& Geen}{He
  et~al.}{2020}]{He2020}
He C.-C.,  Ricotti M.,   Geen S.,  2020, \mn@doi [Monthly Notices of the Royal
  Astronomical Society] {10.1093/mnras/staa165}, 492, 4858

\bibitem[\protect\citeauthoryear{Hollyhead, Bastian, Adamo, Silva-Villa, Dale,
  Ryon  \& Gazak}{Hollyhead et~al.}{2015}]{hollyhead2015}
Hollyhead K.,  Bastian N.,  Adamo A.,  Silva-Villa E.,  Dale J.,  Ryon J.~E.,
  Gazak Z.,  2015, \mn@doi [Monthly Notices of the Royal Astronomical Society]
  {10.1093/mnras/stv331}, 449, 1106

\bibitem[\protect\citeauthoryear{Hui \& Gnedin}{Hui \&
  Gnedin}{1997}]{hui_equation_1997}
Hui L.,  Gnedin N.~Y.,  1997, \mn@doi [Mon Not R Astron Soc]
  {10.1093/mnras/292.1.27}, 292, 27

\bibitem[\protect\citeauthoryear{Hunter}{Hunter}{2007}]{Hunter:2007}
Hunter J.~D.,  2007, \mn@doi [Computing in Science \& Engineering]
  {10.1109/MCSE.2007.55}, 9, 90

\bibitem[\protect\citeauthoryear{Iliev, Mellema, Ahn, Shapiro, Mao  \&
  Pen}{Iliev et~al.}{2014}]{iliev_simulating_2014}
Iliev I.~T.,  Mellema G.,  Ahn K.,  Shapiro P.~R.,  Mao Y.,   Pen U.-L.,  2014,
  \mn@doi [Mon Not R Astron Soc] {10.1093/mnras/stt2497}, 439, 725

\bibitem[\protect\citeauthoryear{Ishigaki, Kawamata, Ouchi, Oguri, Shimasaku
  \& Ono}{Ishigaki et~al.}{2018}]{ishigaki_full-data_2018}
Ishigaki M.,  Kawamata R.,  Ouchi M.,  Oguri M.,  Shimasaku K.,   Ono Y.,
  2018, \mn@doi [The Astrophysical Journal] {10.3847/1538-4357/aaa544}, 854, 73

\bibitem[\protect\citeauthoryear{{Kannan}, {Garaldi}, {Smith}, {Pakmor},
  {Springel}, {Vogelsberger}  \& {Hernquist}}{{Kannan}
  et~al.}{2022}]{kannan_introducing_2021}
{Kannan} R.,  {Garaldi} E.,  {Smith} A.,  {Pakmor} R.,  {Springel} V.,
  {Vogelsberger} M.,   {Hernquist} L.,  2022, \mn@doi [\mnras]
  {10.1093/mnras/stab3710}, \href
  {https://ui.adsabs.harvard.edu/abs/2022MNRAS.511.4005K} {511, 4005}

\bibitem[\protect\citeauthoryear{{Katz}}{{Katz}}{2022}]{katz_non-equilibrium_2022}
{Katz} H.,  2022, \mn@doi [\mnras] {10.1093/mnras/stac423}, \href
  {https://ui.adsabs.harvard.edu/abs/2022MNRAS.512..348K} {512, 348}

\bibitem[\protect\citeauthoryear{Katz, Kimm, Sijacki  \& Haehnelt}{Katz
  et~al.}{2017}]{katz_interpreting_2017}
Katz H.,  Kimm T.,  Sijacki D.,   Haehnelt M.~G.,  2017, \mn@doi [Mon Not R
  Astron Soc] {10.1093/mnras/stx608}, 468, 4831

\bibitem[\protect\citeauthoryear{Kennicutt}{Kennicutt}{1998}]{kennicutt_star_1998}
Kennicutt Jr. R.~C.,  1998, \mn@doi [Annual Review of Astronomy and
  Astrophysics] {10.1146/annurev.astro.36.1.189}, 36, 189

\bibitem[\protect\citeauthoryear{Khusanova et~al.,}{Khusanova
  et~al.}{2020}]{khusanova_alpine-alma_2020}
Khusanova Y.,  et~al., 2020, arXiv e-prints, 2007, arXiv:2007.08384

\bibitem[\protect\citeauthoryear{Kimm \& Cen}{Kimm \&
  Cen}{2014}]{kimm_escape_2014}
Kimm T.,  Cen R.,  2014, \mn@doi [The Astrophysical Journal]
  {10.1088/0004-637X/788/2/121}, 788, 121

\bibitem[\protect\citeauthoryear{Kirby, Cohen, Guhathakurta, Cheng, Bullock  \&
  Gallazzi}{Kirby et~al.}{2013}]{kirby_universal_2013}
Kirby E.~N.,  Cohen J.~G.,  Guhathakurta P.,  Cheng L.,  Bullock J.~S.,
  Gallazzi A.,  2013, \mn@doi [The Astrophysical Journal]
  {10.1088/0004-637X/779/2/102}, 779, 102

\bibitem[\protect\citeauthoryear{Laporte et~al.,}{Laporte
  et~al.}{2017}]{laporte_dust_2017}
Laporte N.,  et~al., 2017, \mn@doi [The Astrophysical Journal Letters]
  {10.3847/2041-8213/aa62aa}, 837, L21

\bibitem[\protect\citeauthoryear{Levermore}{Levermore}{1984}]{levermore_relating_1984}
Levermore C.~D.,  1984, \mn@doi [Journal of Quantitative Spectroscopy and
  Radiative Transfer] {10.1016/0022-4073(84)90112-2}, 31, 149

\bibitem[\protect\citeauthoryear{Lewis et~al.,}{Lewis
  et~al.}{2020}]{lewis_galactic_2020}
Lewis J. S.~W.,  et~al., 2020, \mn@doi [Monthly Notices of the Royal
  Astronomical Society] {10.1093/mnras/staa1748}

\bibitem[\protect\citeauthoryear{Livermore, Finkelstein  \& Lotz}{Livermore
  et~al.}{2017}]{livermore_directly_2017}
Livermore R.~C.,  Finkelstein S.~L.,   Lotz J.~M.,  2017, \mn@doi [The
  Astrophysical Journal] {10.3847/1538-4357/835/2/113}, 835, 113

\bibitem[\protect\citeauthoryear{{Lovell}, {Vijayan}, {Thomas}, {Wilkins},
  {Barnes}, {Irodotou}  \& {Roper}}{{Lovell} et~al.}{2021}]{lovell_first_2021}
{Lovell} C.~C.,  {Vijayan} A.~P.,  {Thomas} P.~A.,  {Wilkins} S.~M.,  {Barnes}
  D.~J.,  {Irodotou} D.,   {Roper} W.,  2021, \mn@doi [\mnras]
  {10.1093/mnras/staa3360}, \href
  {https://ui.adsabs.harvard.edu/abs/2021MNRAS.500.2127L} {500, 2127}

\bibitem[\protect\citeauthoryear{Ma et~al.,}{Ma
  et~al.}{2018}]{ma_simulating_2018}
Ma X.,  et~al., 2018, \mn@doi [Monthly Notices of the Royal Astronomical
  Society] {10.1093/mnras/sty1024}, 478, 1694

\bibitem[\protect\citeauthoryear{Madau, Pozzetti  \& Dickinson}{Madau
  et~al.}{1998}]{madau_star_1998}
Madau P.,  Pozzetti L.,   Dickinson M.,  1998, \mn@doi [ApJ] {10.1086/305523},
  498, 106

\bibitem[\protect\citeauthoryear{Mancini, Schneider, Graziani, Valiante, Dayal,
  Maio, Ciardi  \& Hunt}{Mancini et~al.}{2015}]{mancini_dust_2015}
Mancini M.,  Schneider R.,  Graziani L.,  Valiante R.,  Dayal P.,  Maio U.,
  Ciardi B.,   Hunt L.~K.,  2015, \mn@doi [Monthly Notices of the Royal
  Astronomical Society] {10.1093/mnrasl/slv070}, 451, L70

\bibitem[\protect\citeauthoryear{Maselli, Ferrara  \& Ciardi}{Maselli
  et~al.}{2003}]{maselli_crash_2003}
Maselli A.,  Ferrara A.,   Ciardi B.,  2003, \mn@doi [Monthly Notices of the
  Royal Astronomical Society] {10.1046/j.1365-8711.2003.06979.x}, 345, 379

\bibitem[\protect\citeauthoryear{McKee}{McKee}{1989}]{mckee_photoionization_1989}
McKee C.~F.,  1989, \mn@doi [The Astrophysical Journal] {10.1086/167950}, 345,
  782

\bibitem[\protect\citeauthoryear{Novak, Ostriker  \& Ciotti}{Novak
  et~al.}{2012}]{novak_radiative_2012}
Novak G.~S.,  Ostriker J.~P.,   Ciotti L.,  2012, \mn@doi [Monthly Notices of
  the Royal Astronomical Society] {10.1111/j.1365-2966.2012.21844.x}, 427, 2734

\bibitem[\protect\citeauthoryear{Ocvirk et~al.,}{Ocvirk
  et~al.}{2016}]{ocvirk_cosmic_2016}
Ocvirk P.,  et~al., 2016, \mn@doi [Monthly Notices of the Royal Astronomical
  Society] {10.1093/mnras/stw2036}, 463, 1462

\bibitem[\protect\citeauthoryear{Ocvirk, Aubert, Chardin, Deparis  \&
  Lewis}{Ocvirk et~al.}{2019}]{ocvirk_impact_2019}
Ocvirk P.,  Aubert D.,  Chardin J.,  Deparis N.,   Lewis J.,  2019, \mn@doi
  [Astronomy and Astrophysics] {10.1051/0004-6361/201832923}, 626, A77

\bibitem[\protect\citeauthoryear{Ocvirk et~al.,}{Ocvirk
  et~al.}{2020}]{ocvirk_cosmic_2020}
Ocvirk P.,  et~al., 2020, \mn@doi [Monthly Notices of the Royal Astronomical
  Society] {10.1093/mnras/staa1266}

\bibitem[\protect\citeauthoryear{Ocvirk, Lewis, Gillet, Chardin, Aubert,
  Deparis  \& Thelie}{Ocvirk et~al.}{2021}]{ocvirk_lyman-alpha_2021}
Ocvirk P.,  Lewis J. S.~W.,  Gillet N.,  Chardin J.,  Aubert D.,  Deparis N.,
  Thelie E.,  2021, arXiv e-prints, 2105, arXiv:2105.01663

\bibitem[\protect\citeauthoryear{Oesch et~al.,}{Oesch
  et~al.}{2018}]{oesch_hduv_2018}
Oesch P.~A.,  et~al., 2018, \mn@doi [The Astrophysical Journal Supplement
  Series] {10.3847/1538-4365/aacb30}, 237, 12

\bibitem[\protect\citeauthoryear{Park et~al.,}{Park
  et~al.}{2021}]{park_crucial_2021}
Park H.,  et~al., 2021, arXiv e-prints, p. arXiv:2105.10770

\bibitem[\protect\citeauthoryear{Pawlik, Rahmati, Schaye, Jeon  \&
  Dalla~Vecchia}{Pawlik et~al.}{2017}]{pawlik_aurora_2017}
Pawlik A.~H.,  Rahmati A.,  Schaye J.,  Jeon M.,   Dalla~Vecchia C.,  2017,
  \mn@doi [Monthly Notices of the Royal Astronomical Society]
  {10.1093/mnras/stw2869}, 466, 960

\bibitem[\protect\citeauthoryear{{Planck Collaboration} et~al.,}{{Planck
  Collaboration} et~al.}{2018}]{planck_collaboration_planck_2018}
{Planck Collaboration} et~al., 2018, arXiv e-prints, 1807, arXiv:1807.06205

\bibitem[\protect\citeauthoryear{Popping, Somerville  \& Galametz}{Popping
  et~al.}{2017}]{popping_dust_2017}
Popping G.,  Somerville R.~S.,   Galametz M.,  2017, \mn@doi [Monthly Notices
  of the Royal Astronomical Society] {10.1093/mnras/stx1545}, 471, 3152

\bibitem[\protect\citeauthoryear{{Pozzi}, {Calura}, {Zamorani}, {Delvecchio},
  {Gruppioni}  \& {Santini}}{{Pozzi} et~al.}{2020}]{pozzi_dust_2020}
{Pozzi} F.,  {Calura} F.,  {Zamorani} G.,  {Delvecchio} I.,  {Gruppioni} C.,
  {Santini} P.,  2020, \mn@doi [\mnras] {10.1093/mnras/stz2724}, \href
  {https://ui.adsabs.harvard.edu/abs/2020MNRAS.491.5073P} {491, 5073}

\bibitem[\protect\citeauthoryear{Prunet \& Pichon}{Prunet \&
  Pichon}{2013}]{prunet_mpgrafic_2013}
Prunet S.,  Pichon C.,  2013, Astrophysics Source Code Library, p.
  ascl:1304.014

\bibitem[\protect\citeauthoryear{Puchwein, Haardt, Haehnelt  \& Madau}{Puchwein
  et~al.}{2019}]{puchwein_consistent_2019}
Puchwein E.,  Haardt F.,  Haehnelt M.~G.,   Madau P.,  2019, \mn@doi [Monthly
  Notices of the Royal Astronomical Society] {10.1093/mnras/stz222}

\bibitem[\protect\citeauthoryear{Rasera \& Teyssier}{Rasera \&
  Teyssier}{2006}]{rasera_history_2006}
Rasera Y.,  Teyssier R.,  2006, \mn@doi [Astronomy and Astrophysics]
  {10.1051/0004-6361:20053116}, 445, 1

\bibitem[\protect\citeauthoryear{Read, Iorio, Agertz  \& Fraternali}{Read
  et~al.}{2017}]{read_stellar_2017}
Read J.~I.,  Iorio G.,  Agertz O.,   Fraternali F.,  2017, \mn@doi [Monthly
  Notices of the Royal Astronomical Society] {10.1093/mnras/stx147}, 467, 2019

\bibitem[\protect\citeauthoryear{{R{\'e}my-Ruyer} et~al.,}{{R{\'e}my-Ruyer}
  et~al.}{2015}]{ruyer_linking_2015}
{R{\'e}my-Ruyer} A.,  et~al., 2015, \mn@doi [\aap]
  {10.1051/0004-6361/201526067}, \href
  {https://ui.adsabs.harvard.edu/abs/2015A&A...582A.121R} {582, A121}

\bibitem[\protect\citeauthoryear{Rosdahl, Blaizot, Aubert, Stranex  \&
  Teyssier}{Rosdahl et~al.}{2013}]{rosdahl_ramses-rt:_2013}
Rosdahl J.,  Blaizot J.,  Aubert D.,  Stranex T.,   Teyssier R.,  2013, \mn@doi
  [Monthly Notices of the Royal Astronomical Society] {10.1093/mnras/stt1722},
  436, 2188

\bibitem[\protect\citeauthoryear{Rosdahl et~al.,}{Rosdahl
  et~al.}{2018}]{rosdahl_sphinx_2018}
Rosdahl J.,  et~al., 2018, \mn@doi [Monthly Notices of the Royal Astronomical
  Society] {10.1093/mnras/sty1655}, 479, 994

\bibitem[\protect\citeauthoryear{Schaerer, Boone, Zamojski, Staguhn,
  Dessauges-Zavadsky, Finkelstein  \& Combes}{Schaerer
  et~al.}{2015}]{schaerer_new_2015}
Schaerer D.,  Boone F.,  Zamojski M.,  Staguhn J.,  Dessauges-Zavadsky M.,
  Finkelstein S.,   Combes F.,  2015, \mn@doi [A\&A]
  {10.1051/0004-6361/201424649}, 574, A19

\bibitem[\protect\citeauthoryear{Shen et~al.,}{Shen
  et~al.}{2020}]{shen_high-redshift_2020}
Shen X.,  et~al., 2020, \mn@doi [Monthly Notices of the Royal Astronomical
  Society] {10.1093/mnras/staa1423}, 495, 4747

\bibitem[\protect\citeauthoryear{Stefanon, Bouwens, Labbé, Illingworth,
  Gonzalez  \& Oesch}{Stefanon et~al.}{2021}]{stefanon_galaxy_2021}
Stefanon M.,  Bouwens R.~J.,  Labbé I.,  Illingworth G.~D.,  Gonzalez V.,
  Oesch P.~A.,  2021, arXiv e-prints, 2103, arXiv:2103.16571

\bibitem[\protect\citeauthoryear{Teyssier}{Teyssier}{2002}]{teyssier_cosmological_2002}
Teyssier R.,  2002, \mn@doi [A\&A] {10.1051/0004-6361:20011817}, 385, 337

\bibitem[\protect\citeauthoryear{Teyssier, Fromang  \& Dormy}{Teyssier
  et~al.}{2006}]{teyssier_kinematic_2006}
Teyssier R.,  Fromang S.,   Dormy E.,  2006, \mn@doi [Journal of Computational
  Physics] {10.1016/j.jcp.2006.01.042}, 218, 44

\bibitem[\protect\citeauthoryear{Toro, Spruce  \& Speares}{Toro
  et~al.}{1994}]{toro_restoration_1994}
Toro E.~F.,  Spruce M.,   Speares W.,  1994, \mn@doi [Shock Waves]
  {10.1007/BF01414629}, 4, 25

\bibitem[\protect\citeauthoryear{Torrey et~al.,}{Torrey
  et~al.}{2019}]{torrey_evolution_2019}
Torrey P.,  et~al., 2019, \mn@doi [Monthly Notices of the Royal Astronomical
  Society] {10.1093/mnras/stz243}, 484, 5587

\bibitem[\protect\citeauthoryear{Trebitsch et~al.,}{Trebitsch
  et~al.}{2020}]{trebitsch_obelisk_2020}
Trebitsch M.,  et~al., 2020, arXiv e-prints, 2002, arXiv:2002.04045

\bibitem[\protect\citeauthoryear{Van Der~Walt, Colbert  \& Varoquaux}{Van
  Der~Walt et~al.}{2011}]{van2011numpy}
Van Der~Walt S.,  Colbert S.~C.,   Varoquaux G.,  2011, Computing in Science \&
  Engineering, 13, 22

\bibitem[\protect\citeauthoryear{Vijayan, Clay, Thomas, Yates, Wilkins  \&
  Henriques}{Vijayan et~al.}{2019}]{vijayan_detailed_2019}
Vijayan A.~P.,  Clay S.~J.,  Thomas P.~A.,  Yates R.~M.,  Wilkins S.~M.,
  Henriques B.~M.,  2019, \mn@doi [Monthly Notices of the Royal Astronomical
  Society] {10.1093/mnras/stz1948}, p. stz1948

\bibitem[\protect\citeauthoryear{Vijayan, Lovell, Wilkins, Thomas, Barnes,
  Irodotou, Kuusisto  \& Roper}{Vijayan et~al.}{2020}]{vijayan_first_2020}
Vijayan A.~P.,  Lovell C.~C.,  Wilkins S.~M.,  Thomas P.~A.,  Barnes D.~J.,
  Irodotou D.,  Kuusisto J.,   Roper W.,  2020, arXiv:2008.06057 [astro-ph]

\bibitem[\protect\citeauthoryear{{Virtanen} et~al.,}{{Virtanen}
  et~al.}{2020}]{Virtanen_2020}
{Virtanen} P.,  et~al., 2020, \mn@doi [Nature Methods]
  {https://doi.org/10.1038/s41592-019-0686-2}, \href {https://rdcu.be/b08Wh}
  {17, 261}

\bibitem[\protect\citeauthoryear{Wall, Mac~Low, McMillan, Klessen,
  Portegies~Zwart  \& Pellegrino}{Wall et~al.}{2020}]{wall2020}
Wall J.~E.,  Mac~Low M.-M.,  McMillan S. L.~W.,  Klessen R.~S.,
  Portegies~Zwart S.,   Pellegrino A.,  2020, \mn@doi [The Astrophysical
  Journal] {10.3847/1538-4357/abc011}, 904, 192

\bibitem[\protect\citeauthoryear{{Wilkins} et~al.,}{{Wilkins}
  et~al.}{2013}]{wilkins2013_dust_colors}
{Wilkins} S.~M.,  et~al., 2013, \mn@doi [\mnras] {10.1093/mnras/stt1471}, \href
  {https://ui.adsabs.harvard.edu/abs/2013MNRAS.435.2885W} {435, 2885}

\bibitem[\protect\citeauthoryear{{Wilkins}, {Feng}, {Di-Matteo}, {Croft},
  {Stanway}, {Bunker}, {Waters}  \& {Lovell}}{{Wilkins}
  et~al.}{2016}]{wilkins_photon_2016}
{Wilkins} S.~M.,  {Feng} Y.,  {Di-Matteo} T.,  {Croft} R.,  {Stanway} E.~R.,
  {Bunker} A.,  {Waters} D.,   {Lovell} C.,  2016, \mn@doi [\mnras]
  {10.1093/mnras/stw1154}, \href
  {https://ui.adsabs.harvard.edu/abs/2016MNRAS.460.3170W} {460, 3170}

\bibitem[\protect\citeauthoryear{{Wilkins}, {Feng}, {Di Matteo}, {Croft},
  {Lovell}  \& {Waters}}{{Wilkins} et~al.}{2017}]{wilkins_properties_2017}
{Wilkins} S.~M.,  {Feng} Y.,  {Di Matteo} T.,  {Croft} R.,  {Lovell} C.~C.,
  {Waters} D.,  2017, \mn@doi [\mnras] {10.1093/mnras/stx841}, \href
  {https://ui.adsabs.harvard.edu/abs/2017MNRAS.469.2517W} {469, 2517}

\bibitem[\protect\citeauthoryear{Wu, Davé, Tacchella  \& Lotz}{Wu
  et~al.}{2020}]{wu_photometric_2020}
Wu X.,  Davé R.,  Tacchella S.,   Lotz J.,  2020, \mn@doi [Monthly Notices of
  the Royal Astronomical Society] {10.1093/mnras/staa1044}, p. staa1044

\bibitem[\protect\citeauthoryear{Yajima, Choi  \& Nagamine}{Yajima
  et~al.}{2011}]{yajima_escape_2011}
Yajima H.,  Choi J.-H.,   Nagamine K.,  2011, \mn@doi [Mon Not R Astron Soc]
  {10.1111/j.1365-2966.2010.17920.x}, 412, 411

\bibitem[\protect\citeauthoryear{Yoo, Kimm  \& Rosdahl}{Yoo
  et~al.}{2020}]{yoo_origin_2020}
Yoo T.,  Kimm T.,   Rosdahl J.,  2020, arXiv e-prints, 2001, arXiv:2001.05508

\bibitem[\protect\citeauthoryear{{Yung}, {Somerville}, {Popping},
  {Finkelstein}, {Ferguson}  \& {Dav{\'e}}}{{Yung}
  et~al.}{2019}]{yung_semi_2019}
{Yung} L.~Y.~A.,  {Somerville} R.~S.,  {Popping} G.,  {Finkelstein} S.~L.,
  {Ferguson} H.~C.,   {Dav{\'e}} R.,  2019, \mn@doi [\mnras]
  {10.1093/mnras/stz2755}, \href
  {https://ui.adsabs.harvard.edu/abs/2019MNRAS.490.2855Y} {490, 2855}

\bibitem[\protect\citeauthoryear{Zavala et~al.,}{Zavala
  et~al.}{2021}]{zavala_evolution_2021}
Zavala J.~A.,  et~al., 2021, eprint arXiv:2101.04734, p. arXiv:2101.04734

\makeatother
\end{thebibliography}



\appendix

\section{Impact of the dust extinction curve}

\label{app:kappa_pick}



\begin{figure*}
    \includegraphics[trim={0cm 0cm 0.cm 0cm},clip,width=\columnwidth]{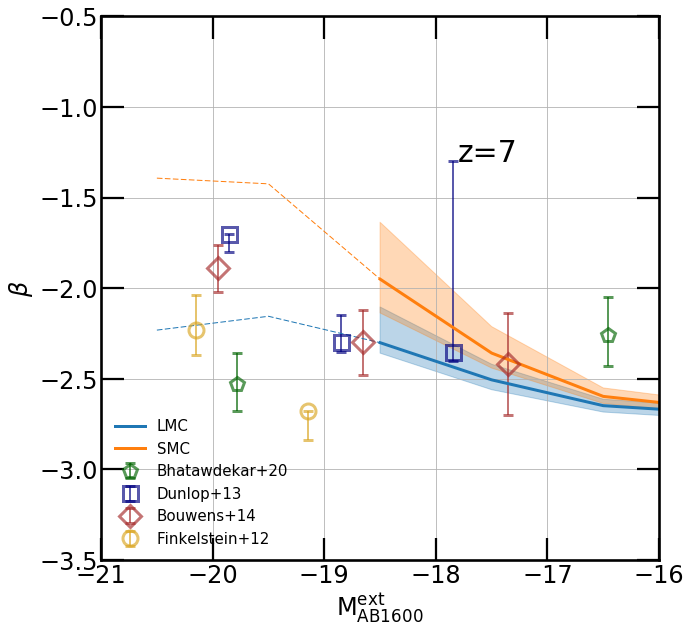}
    \includegraphics[trim={0cm 0cm 0.cm 0cm},clip,width=\columnwidth]{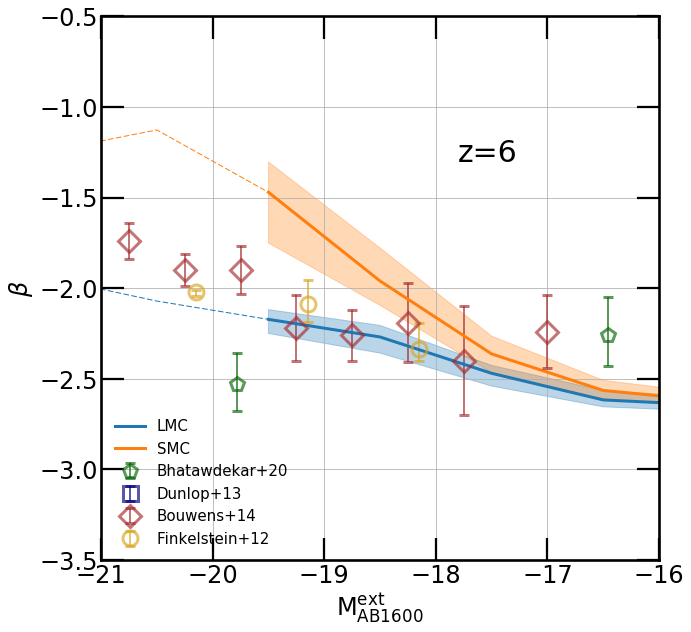}
    \includegraphics[trim={0cm 0cm 0.cm 0cm},clip,width=\columnwidth]{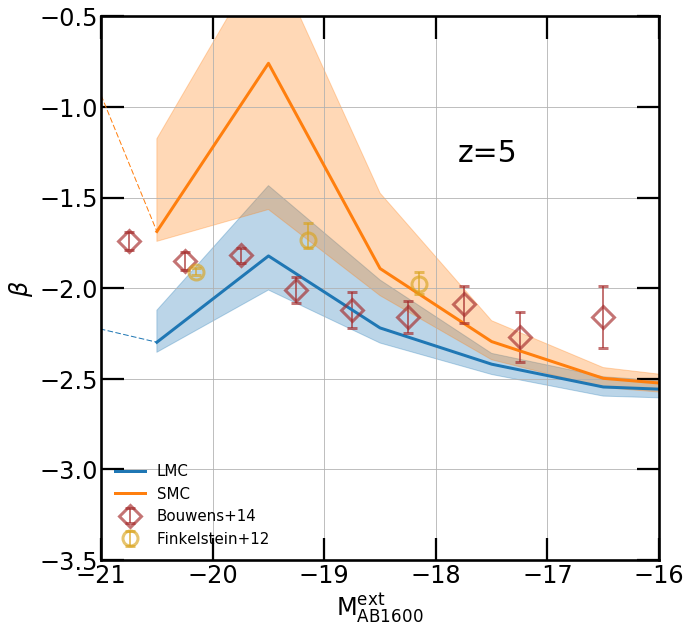}
    \caption{Comparison between the median UV slope $\beta$ versus \magy \, in \dustier{} at $\rm z=7,6,5$ for both LMC based and SMC based extinction curves from \protect \cite{draine_infrared_2001}. The full lines represent the median in bins where there are more than 5 galaxies, whereas the thin lines represent the median in bins where there are less than or exactly 5 galaxies. The shaded areas show the 16\% and 84\% $\beta$ values for each \magy \, bin. Various observational constraints are shown \protect \citep[][]{bouwens_census_2014,dunlop_uv_2013,finkelstein_candels_2012,bhatawdekar_uv_2020}. To produce the points from \protect \cite{finkelstein_candels_2012}, we took the observed data and processed them as we did for the simulation data.}
    \label{fig:SMCvsLMC}
    \end{figure*}

The choice of the dust extinction curve $\rm \kappa_d$ has potentially significant repercussions on the final values of the UV slope $\beta$, the luminosity function, and the fraction of dust-obscured star formation $\rm f_{obs}$. Fig. \ref{fig:SMCvsLMC} illustrates this by showing the average $\beta$ as a function of \magy \, in \dustier{} haloes for the two best matching extinction curves we tested, at several redshifts when compared to observational constraints. Whether using a LMC or SMC extinction curve, the median $\beta$ is redder for brighter \magy. However the SMC extinction curve leads to a much steeper rise of the UV slope $\beta$ with increasing brightness than the LMC ones. The two median curves are very similar for \magy$\geq-17.5$. But from this point onwards, the brighter the magnitude, the greater the gap between the SMC based result and the LMC based one. By $\rm z=5$ and at \magy=-19.5, the SMC extinction curve predicts a median $\beta$ of roughly -0.75 whereas the LMC curve predicts a bluer median just above -1.8. As noted in Fig. \ref{fig:betas}, the scatter around the median $\beta$ values is large (particularly for the SMC curves), and increases towards brighter magnitudes. For both extinction curves, there is some evolution with redshift for the brightest galaxies (\magy$<-19$).

{At faint magnitudes, neither of the extinction curves provides a perfect match to constraints at all redshifts, with both sets of curves predicting bluer (<0.2) UV slopes. However, for brighter galaxies, the LMC curves predict much more moderate reddening, leading to a far better match to the constraints for \dustier{}'s bright galaxies. As highlighted in Sec. \ref{sec:pred}, our comparison to observations could be compromised by the small sample of bright galaxies in \dustier{}.
}
There may be room for improvement between our predictions and the observations presented in this appendix, either through an even finer tuning of the extinction curve, beyond the usual SMC/LMC/MW trinity, or of the parameters of the dust model, or both. It is also likely that they are degenerate to some degree, and that forcing them to produce galaxies closer to the observational points would not necessarily improve our understanding of the underlying physics. Therefore we consider our current model using the LMC extinction curve good enough for the goal we set ourselves for this study.

\begin{figure*}
    \centering
    \includegraphics[width=\columnwidth]{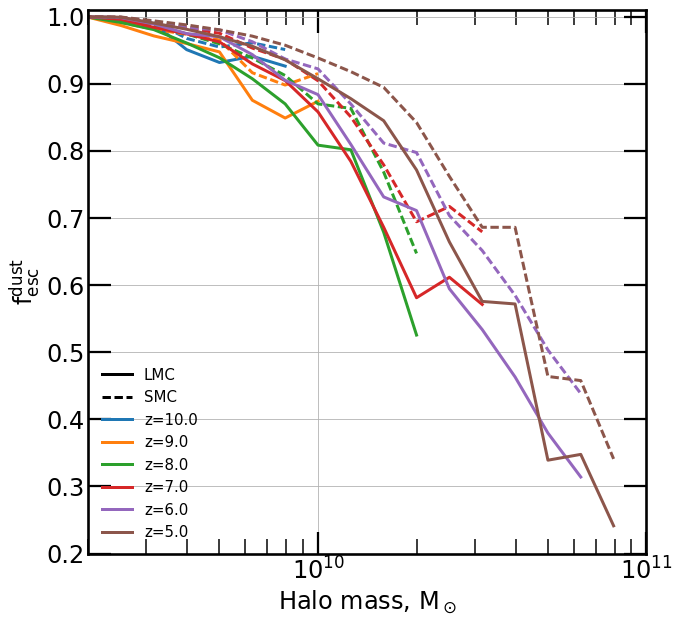}
    \includegraphics[width=\columnwidth]{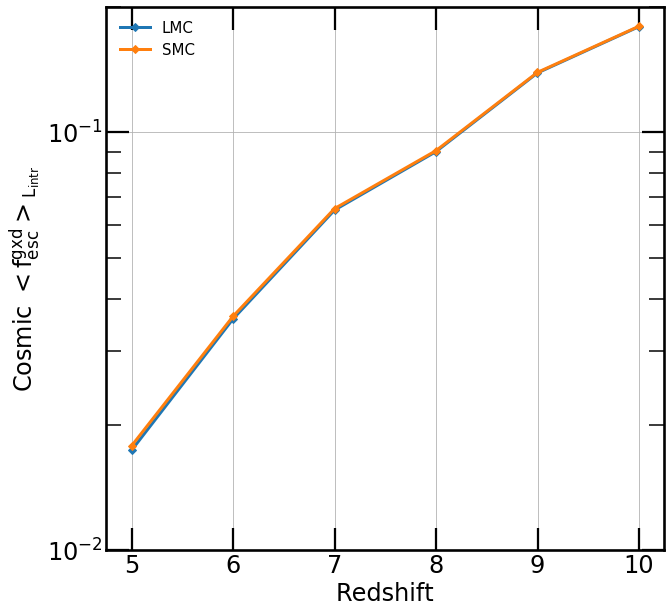}
    \caption{Comparison between \fescd{} results when assuming either an SMC or LMC extinction law. \emph{Left:} The median \fescd{} in \dustier{} galaxies in both cases. \emph{Right:} The luminosity weighted cosmic escape fraction for both extinction laws. Though the different extinction laws give rise to fairly different median \fescd{} for the most massive haloes, the overall average fraction of ionizing photons that escape to the IGM is hardly affected.}
    \label{fig:SMCvsLMC_fesc}
\end{figure*}

{The chosen extinction law also affects the transfer of ionizing photons. Although we find LMC values lead to more realistic reddening of the UV slopes of galaxies, SMC extinction values were used at run time for the radiative transfer of photons. In the rest of our post-processing, the escape of ionizing photons is computed using an LMC extinction law throughout the paper, introducing an inconsistency with the  run-time assumptions. In the left panel of Fig. \ref{fig:SMCvsLMC_fesc}, we show the median \fescd{} for the LMC and SMC extinction laws. For $\rm M_h<$\SI{2e9}{\simsun} \fescd{}$\approx 1.0$ for both extinction laws. For higher masses, the LMC \fescd{} values are consistently lower than their SMC based counterparts. The greater the halo mass, and the lower the redshift, the larger the step between the two medians. The difference between the two results for the highest mass haloes at $\rm z=6$ is roughly 0.1. Indeed, the LMC extinction law makes the dust optical depths approximately $1.5 \times$ higher than the SMC equivalents. Though this difference may seem significant for the high mass haloes, the right panel of Fig. \ref{fig:SMCvsLMC_fesc} shows that the overall impact on the average fraction of ionizing photons that escape to the IGM is very slight, even at $\rm z=5$. This is because the total escape of ionizing photons remains dominated by absorption from galactic neutral Hydrogen. Thus, our overall conclusion of a modest effect of dust would assuredly not be significantly affected by running a simulation with an LMC extinction curve. However, the left panel does suggest that there could be some differences in the growth of the largest HII bubbles, as well as in the ionizing radiation seen by the satellite galaxies surrounding the most massive haloes in the simulations (potentially affecting radiative suppression in the massive galaxies' satellites).}

\section{Numerical implementation of dust in ATON's radiative transfer}

\label{app:dust_rt}

In order to account for dust absorption in the radiative transfer module ATON, we must update the radiative transfer equations as done in Sec. 3.3 of \citet{aubert_radiative_2008}.
Starting from the set of transport equations (Eq. 32,33 in \citet{aubert_radiative_2008}), we add a dust absorption term that is analogous to the neutral Hydrogen one, and obtain Eq. \ref{eq:transfer_dust}.

\begin{equation}
    \left\{
\begin{array}{lll}
  \rm \frac{dN_\gamma}{dt} & = & \rm n_e n_{HII}(\alpha_A-\alpha_B) - n_{HI} \sigma_\gamma c N_\gamma - \rho_d\kappa_d c N_\gamma\, , \\ 
  \\
  \rm \frac{d\mathbf{F}_\gamma}{dt} & = & \rm - n_{HI} \sigma_\gamma c \rm\mathbf{F}_\gamma - \rho_d\kappa_d c \rm\mathbf{F}_\gamma\, ,
\end{array}
\right.
\label{eq:transfer_dust}
\end{equation}
here, $\rm n_e$ is the electron density, $\rm n_{\rm HII}$ is the proton density, $\rm \alpha_A$ is the case A recombination coefficient, $\rm \alpha_B$ is the case B recombination coefficient, $\rm n_{\rm HI}$ is the neutral Hydrogen density, $\rm \sigma_\gamma=2.93055 \cdot 10^{-22} m^{2}$ is the effective ionising cross section for 20.28 eV photons, $\rm N_\gamma$ is the ionising photon density, $\rm \mathbf{F_\gamma}$ is the ionising photon flux, and c the speed of light, $\rm \rho_d$ is the dust mass density in a cell in \si{\gram\per\metre\cubed}, and $\rm \kappa_{d} = $\SI{8.85}{\meter\squared\per\gram} is the dust mass attenuation coefficient  at $\rm \lambda=611\mathang$ (roughly the wavelength that corresponds to the energy of 20.28 eV photons) derived for the SMC by \cite{draine_infrared_2001}.

Then we can follow the same steps as in \citet{aubert_radiative_2008} (to reach an equivalent of their Eqs. 36 and 37), and obtain a numerical scheme for updating the number density and flux of photons in a cell in \eqref{eq:transfer_dust_numerical}.

\begin{equation}
    \left\{
\begin{array}{lll}
  \rm  N^{p+1}_\gamma & = & \rm \frac{\Delta t\big(n^{p+1}_e n^{p+1}_{HII} [\alpha^{p+1}_A-\alpha^{p+1}_B]\big) + N^p_\gamma}{1 + c \Delta t  [n^{p+1}_{HI} \sigma_\gamma+\kappa_{d} \rho^{p+1}_d]}\, , \\
  \\
  \rm  \mathbf{F}^{p+1}_\gamma& = & \rm \frac{\mathbf{F}^p_\gamma}{1+ c \Delta t[\sigma_\gamma n^{p+1}_{HI} + \kappa_{d} \rho^{p+1}_d]} \, , \\
\end{array}
\right.
\label{eq:transfer_dust_numerical}
\end{equation}

Where $\rm \Delta_t$ denotes the ATON time step that separates steps p and p+1.

\section{Impact of emission lines on photometric properties and UV slopes}

\label{app:emlines}

\begin{table}
\begin{center}
\begin{tabular}{|c|}
\hline
Species - Wavelength (\AA) \\
\hline
$[$Si II$]$1263\AA \\
$[$Si II$]$1308\AA \\
$[$O I$]$1357\AA \\
$[$Si II$]$1531\AA \\
{\bf $[$C IV$]$1548\AA }\\
{\bf $[$C IV$]$1551\AA }\\
{\bf HeII 1640\AA } \\
$[$O III$]$1661\AA \\
$[$O III$]$1666\AA \\
{\bf $[$C III$]$1907\AA } \\
{\bf $[$C III$]$1910\AA } \\
\hline
\end{tabular}
\end{center}
\caption{List of the chemical species considered in our nebular emission lines models. The brightest lines or systems of lines are noted in bold form.}
\label{t:emlinespecies}
\end{table}

\begin{figure*}
\includegraphics[angle=90,trim={4cm 1cm 2cm 0cm},clip,width=\columnwidth]{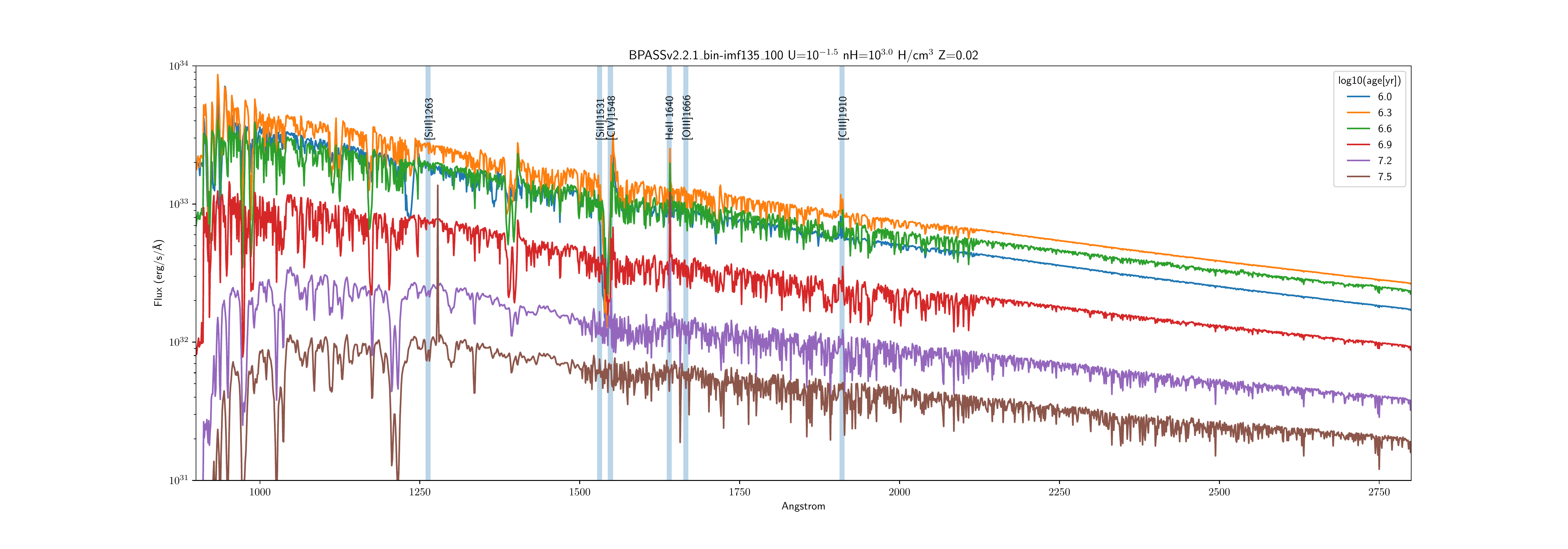}
\includegraphics[angle=90,trim={4cm 1cm 2cm 0cm},clip,width=\columnwidth]{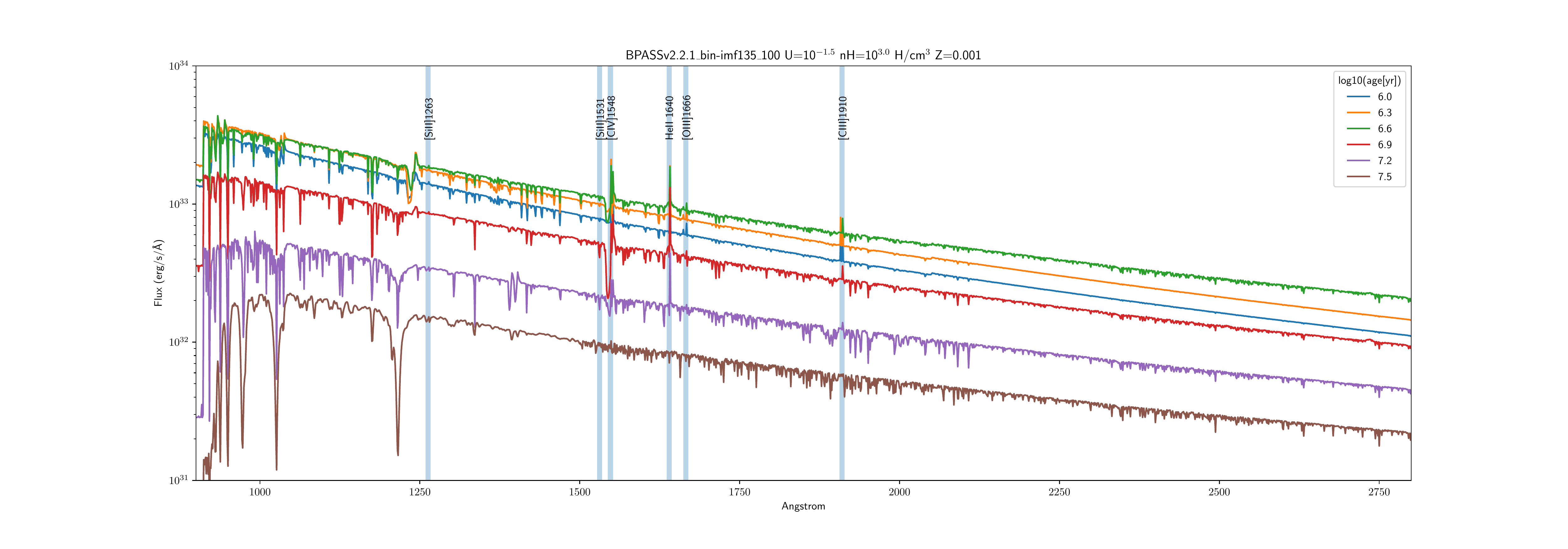}
\caption{BPASS stellar population models spectra for a mass of 1 $\msun$, including the most prominent emission lines predicted, in the 1000-2750 \AA{} range. The ionization parameter U and hydrogen number density have been chosen so as to maximize the strength of the lines in the available models. The figures are provided for metallicities Z=0.02 (i.e. solar) and Z=0.001.}
\label{fig:emlines}
\end{figure*}

\begin{figure}
    \includegraphics[width=1.1\columnwidth]{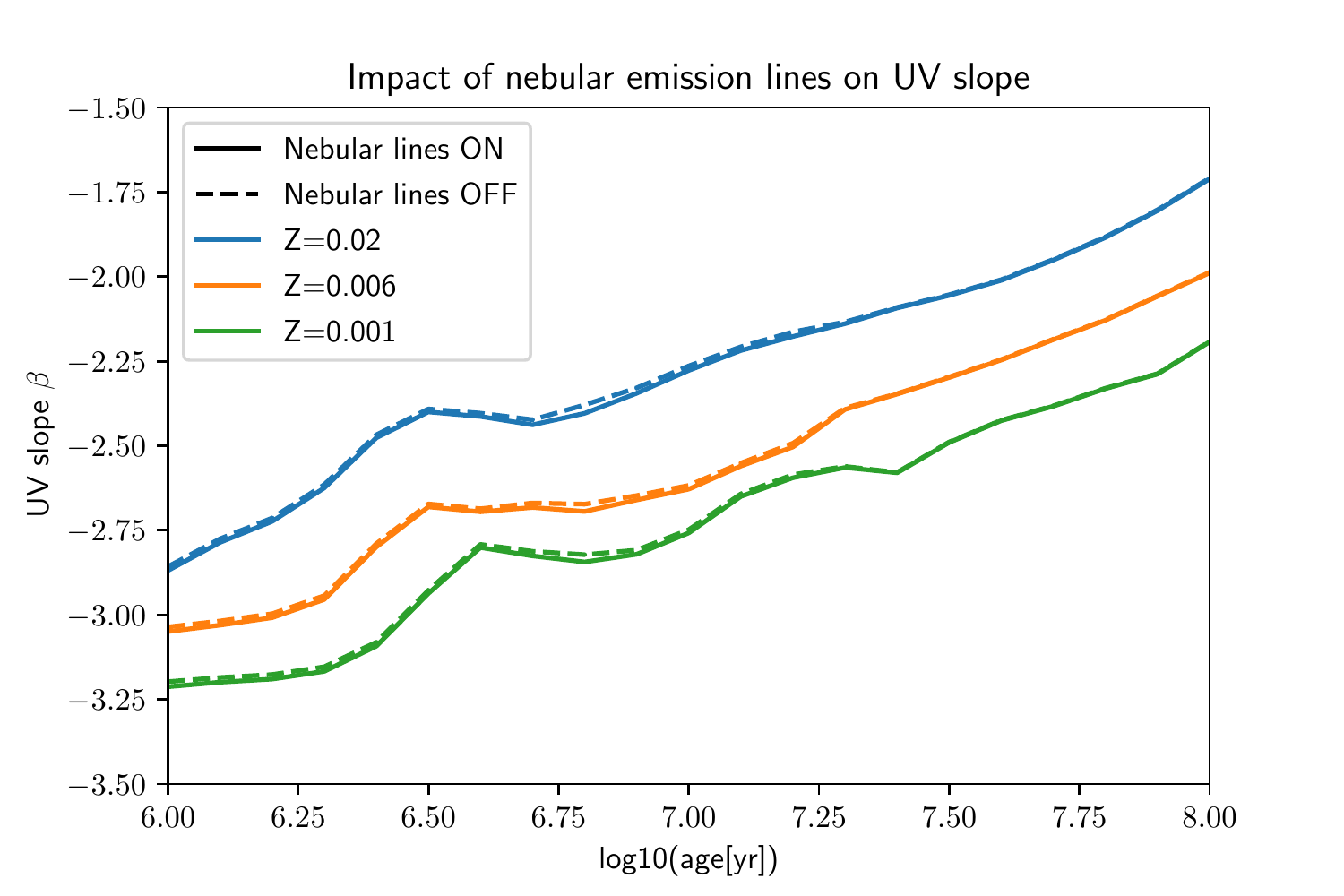}
    \caption{Evolution of UV slopes of BPASS stellar population models as a function of stellar age, for a set of metallicities, and including or neglecting nebular emission lines (solid versus dashed lines).}
    \label{fig:emlinesbetas}
\end{figure}

The prediction of UV slopes and photometric properties of high-redshift galaxies as we propose in this study can be impacted by nebular emission lines falling within camera filters. This is especially true in the infra-red and in the optical, where emission lines can be contribute a large fraction or dominate the measured photometric flux in a given filter. Consequently, colors may also be impacted, as shown in \cite{wilkins2013_dust_colors} for 1500\AA{} - optical colors. Here, we investigate the possible impact of nebular emission lines on the UV spectra of our simulated galaxies and their UV slopes in particular. 
While the BPASS models do not explicitly include nebular emission lines, the authors have made available a set of model nebular emission lines produced using CLOUDY \citep{ferland_cloudy_1998}, for a spherical gas geometry of fixed density, spanning a grid of density, ionization parameter and metallicity. The models are described here \url{https://bpass.auckland.ac.nz/4.html}, and we quickly, for the sake of consistency, reproduce their main properties here. The model cloud is irradiated by an instantaneous burst BPASS stellar population model including binary stars and featuring the same IMF as in the rest of the paper, albeit with a different upper mass limit, set at 300 $\msun$ for these nebular emission lines calculation, while we have used 100 $\msun$ in the rest of the paper. Despite this difference, these models are useful as an upper limit on the emission line fluxes (in the framework considered) as the more massive BPASS IMF produces more ionizing photons.
Emission line fluxes for a given specie are a function of the stellar population age, the ionization parameter U, and the hydrogen gas density. In order to put an upper limit on the expected impact of nebular emission lines, we searched for the ionization parameter and hydrogen gas density which would produce the most prominent emission lines. We found the parameters U=10$^{-1.5}$ and n$_{\rm H}=10^3$ cm$^{-3}$ to be the most conducive to strong emission lines in the grid used, for the species given in Tab. \ref{t:emlinespecies}, and the strongest lines or systems of lines we found in this setup are noted in bold in the table. The Lyman-$\alpha$ line at 1216\AA{}, while extremely bright, is omitted because it falls bluewards of the pseudo-filters we use for computing UV slopes. Besides, its modelling requires accounting for the complex resonant nature of this line, which is beyond the scope of this paper. The resulting spectra, including both stellar continuum and nebular emission lines, are shown in Fig. \ref{fig:emlines} for 1 $\msun$ of stellar population born in an instantaneous burst. The lines are strongest when the stellar population is youngest, i.e. within 10 Myr. 

Most of the strong emission lines are located within 1200-1700 \AA{}, whereas the 2200-2700 \AA{} appears rather line-free.
As a consequence, their presence may impact the UV $\beta$ slope we compute, given that the blue band we use for the slope computation is affected, while the red band is not. In order to quantify this, we computed the slopes of our BPASS stellar populations with and without nebular emission lines, following the procedure described in Sec. \ref{sec:extc}, i.e. using blue and red pseudo-filters. The results are shown in Fig. \ref{fig:emlinesbetas}.
The UV slopes get redder as age increases, and with increasing metallicity, as expected and commonly found in the literature.
Including the emission lines makes the UV slopes bluer by a few per cent, which is very small compared to, e.g. the observational uncertainties on this parameter or the dispersion in our simulated galaxy sample and the impact of e.g. chemical enrichment. Therefore, within the confines of the standard BPASS emission line models used here, the results presented in the main body of the paper are unlikely to be strongly affected by the inclusion of nebular emission lines.



\bsp	
\label{lastpage}
\end{document}